\documentclass[a4paper,fleqn,usenatbib]{mnras}
\usepackage{newtxtext,newtxmath}

\usepackage{graphicx}	
\usepackage{amsmath}	
\usepackage{amssymb}	
\usepackage{multicol}        
\usepackage{bm}		
\usepackage{pdflscape}	

\usepackage[T1]{fontenc}
\usepackage{ae,aecompl}
\usepackage{epstopdf}
\usepackage[usenames, dvipsnames]{color}

\title[prompt GRBs from magnetar engines]{Constraints on millisecond magnetars as the engines of prompt emission in gamma-ray bursts.}
\author[P. Beniamini, D. Giannios, B. D. Metzger]{Paz Beniamini$^{1}$, Dimitrios Giannios$^2$ and Brian D. Metzger$^3$
	\\
	$^1$Department of Physics, The George Washington University, Washington, DC 20052, USA \\
	$^2$Department of Physics and Astronomy, Purdue University, 525 Northwestern Avenue, West Lafayette, IN 47907, USA \\
	$^3$Physics Department and Columbia Astrophysics Laboratory,
	Columbia University, New York, NY 10027, USA}

\begin{document}
	\label{firstpage}
	\pagerange{\pageref{firstpage}--\pageref{lastpage}}
	\maketitle
	
	\begin{abstract}
		We examine millisecond magnetars as central engines of Gamma Ray Bursts' (GRB) prompt emission. Using the proto-magnetar wind model of \cite{Metzger+11}, we estimate the temporal evolution of the magnetization and power injection at the base of the GRB jet and apply these to different prompt emission models to make predictions for the GRB energetics, spectra and lightcurves. We investigate both shock and magnetic reconnection models for the particle acceleration, as well as the effects of energy dissipation across optically thick and thin regions of the jet. The magnetization at the base of the jet, $\sigma_0$, is the main parameter driving the GRB evolution in the magnetar model and the emission is typically released for $100\lesssim \sigma_0 \lesssim 3000$. Given the rapid increase in $\sigma_0$ as the proto-magnetar cools and its neutrino-driven mass loss subsides, the GRB duration is typically limited to $\lesssim 100$ s. This low baryon loading at late times challenges magnetar models for ultra-long GRBs, though black hole models likely run into similar difficulties without substantial entrainment from the jet walls. The maximum radiated gamma-ray energy is $\lesssim 5 \times 10^{51}$erg, significantly less than the magnetar's total initial rotational energy and in strong tension with the high end of the observed GRB energy distribution. However, the gradual magnetic dissipation model \citep{BG2017} applied to a magnetar central engine, naturally explains several key observables of typical GRBs, including energetics, durations, stable peak energies, spectral slopes and a hard to soft evolution during the burst.
	\end{abstract}
	
	
	\section{Introduction}
	
	Long-duration gamma-ray bursts (GRBs) originate from the deaths of massive stars in core collapse supernovae \citep{Woosley93,Stanek+03,Hjorth+03,Woosley&Bloom06}.  However, the nature of the central compact objects responsible for powering their relativistic jets remains a long-standing mystery.  
	
	If the core of a massive rotating star collapses to form a black hole, then accretion of the outer stellar layers through a centrifugally-supported disk \citep{MacFadyen&Woosley99} may supply magnetic flux to the black hole, powering a relativistic jet through the Blandford-Znajek process (e.g.~\citealt{Tchekhovskoy&Giannios15}).  This jet subsequently breaks through the surface of the star, powering the GRB.  This is the classical ``collapsar" model, though many generalize this term to denote the generic association of long GRBs with massive star explosions.
	
	On the other hand, most core collapse events first pass through a proto-neutron star (NS) phase \citep{Burrows&Lattimer86}.  If the progenitor star is indeed rotating at the high required rate prior to collapse, the resulting proto-NS will itself be born rapidly spinning, close to the centrifugal break-up rotation period of $P \approx 1$ ms.  As a result of this rapid rotation, the proto-NS will acquire a strong magnetic dipole field $B_{d} \approx 10^{15}-10^{16}$ G due to the rapid growth of the magneto-rotational instability (e.g., \citealt{Moesta+15}). Acting in concert with rapid rotation, this strong magnetic field may aid the supernova explosion (e.g.~\citealt{Burrows+07}).  
	
	A sufficiently strong explosion will stifle accretion onto the proto-NS \citep{Dessart+08}, preventing its collapse into a black hole and instead leave behind, within the cavity carved behind the expanding supernova ejecta, a rapidly spinning, strongly magnetized NS (a `millisecond magnetar').  Magnetic spin-down of the magnetar over minutes or longer following the explosion provides an alternative mechanism for powering a GRB jet.  In the magnetar model, the burst energy is sourced by the birth rotational energy of the magnetar instead of accretion power \citep{Usov92,Thompson+04,Metzger+07,Metzger+11}.  Although the fraction of the magnetar energy placed into a jet is uncertain \citep{Margalit+17}, the basic mechanism by which the surrounding stellar envelope collimates the magnetized wind into a bipolar jet is qualitatively similar to the black hole case \citep{Uzdensky&MacFadyen07,Komissarov&Barkov07,Bucciantini+07,Bucciantini+08,Bucciantini+09}.
	Newly-formed magnetars have also been invoked as the power sources of fast radio bursts (e.g.~\citealt{Lyubarsky14,Metzger+17,Beloborodov17}) and superluminous supernovae \citep{Kasen&Bildsten10,Woosley10}, though the latter require magnetars with weaker magnetic fields than those required to power classical GRB jets.
	
	Distinguishing black hole and magnetar models for long GRBs is challenging because, unlike most other compact object systems in Nature, GRB central engines are not observable directly due to the high optical depth through the jet.  Past work has focused on either excluding \citep{Cenko+10} or favoring \citep{Mazzali+14} magnetar engines, based on a comparison between the energy of the GRB, its associated afterglow or supernova to the predicted upper limit of $3\times 10^{52}$ erg of the magnetar model, corresponding to the maximum rotational energy of a 1.4$M_{\odot}$ NS.  Making such comparisons is challenging because one must correct for the (uncertain) jet beaming angle to transform isotropic equivalent quantities into true energetics.  Accurate measurements of the kinetic energies of supernovae are likewise hindered by challenges in modeling the spectra of highly-asymmetric explosions (e.g.~\citealt{Dessart+17}). 
	
	The GRB prompt emission should contain useful information for distinguishing black hole from magnetar models.  Magnetars in principle offer the best opportunity to make {\it ab initio} predictions for the GRB emission, because their jet luminosity is related to the well-understood rate of magnetic dipole spindown, while the baryon loading of the jet is controlled by neutrino-driven ablation of matter from the proto-NS as it undergoes Kelvin-Helmholtz cooling following the supernova explosion \citep{Thompson+04,Metzger+07}.  Black hole models, by contrast, currently make no similarly robust predictions because the jet baryon loading is instead set by the uncertain rate of entrainment/diffusion from the jet walls (e.g.~\citealt{Levinson&Eichler03}).
	
	Even given a well-defined prediction for the temporal evolution of the jet luminosity and baryon loading, the site and mechanism of GRB prompt emission remains an area of active research. While internal shocks within an optically-thin environment were long considered the most promising prompt emission model (e.g.~\citealt{Rees&Meszaros94,Kobayashi+97,DM1998,DM2000}), work in recent years has focused on emission from higher optical depths within the jet (\citealt{Thompson94,Ghisellini1999,Meszaros2000,Peer+06,Lazzati2010,Beloborodov2010,Levinson2012,Giannios2012}). Furthermore, internal shocks are only efficient in weakly magnetized jets (e.g., \citealt{Sironi2015}); in the strongly magnetized jet flow expected to be launched by the magnetar, magnetic reconnection provides a more efficient process for powering jet emission (\citealt{Spruit+01}).\footnote{In principle neutrino annihilation at the central engine can result in the direct launching of a fireball, allowing for internal shocks to take place after the bulk acceleration process is complete. However, neutrino annihilation has been shown to be inefficient in accounting for the observed luminosity of GRBs (\citealt{Leng&Giannios14}).} This has led to the exploration of magnetically dominated GRB emission regions, with the dissipation being driven by magnetic reconnection (\citealt{Giannios2008,Zhang2011,Mckinney2012,Beniamini2014,Sironi2015,BG2016,Kagan2016,Begue2017}).
	
	In this paper we explore constraints on the magnetar model for GRBs based on its predictions for the prompt emission.  Our method is to use models for the time evolution of the magnetar jet luminosity and baryon loading in conjunction with prompt emission models to create predictions for the high-energy emission energetics, spectra and lightcurves.  For completeness, we do not limit our study to a specific dissipation/emission model for the GRB.  We investigate both shock and magnetic reconnection models for the particle acceleration as well as the effect of dissipation at both optically thick and thin parts of the jet.
	This work builds on earlier work by \citet{Metzger+11} but takes advantage of recent advances in modeling magnetic dissipation models \citep{BG2017}.  We aim to critically assess various GRB observations which are consistent with, or in tension with, the magnetar model.  
	
	This paper is organized as follows.  In $\S\ref{sec:magnetarevolve}$ we describe the magnetar wind evolution, and in $\S\ref{sec:GRB}$ we describe different models for the prompt emission.  In $\S\ref{sec:results}$ we discuss our results.  In $\S\ref{sec:General}$ we provide an overview of comparison to GRB observables.  In $\S\ref{sec:conclusions}$ we provided a bulleted conclusion.  

	\section{Magnetar Wind Model}
	\label{sec:magnetarevolve}
	
	\subsection{Description of Model}
	We consider the proto-magnetar wind model described by \cite{Metzger+11} as the source of the outflow responsible for feeding the relativistic GRB jet. The power source of the wind is the rotational energy of the magnetar. The energy loss rate, $\dot{E}$ \footnote{Here and elsewhere in the text, all quantities are defined in the magnetar's rest frame, unless explicitly mentioned otherwise.} is dictated by magnetic dipole spin-down in the force-free limit \citep{Spitkovsky06}.\footnote{For simplicity, and because it does not lead to large quantitative errors, we neglect corrections to the spin-down rate that occur during the brief phase at early times when the wind is sufficiently baryon-loaded to be non-relativistic (\citealt{Thompson+04,Metzger+07}).} The mass loss rate, $\dot{M}$ is driven by neutrino heating in the atmosphere above the NS surface, as it undergoes Kelvin-Helmholtz contraction and cooling over the first minute after its formation. At later times ($\gtrsim 100$ s), neutrino heating subsides as the proto-NS becomes transparent to neutrinos \citep{Pons+99}.  The magnetosphere thus becomes sufficiently tenuous for vacuum electric fields to develop, which initiates pair creation and causes the wind mass loss rate to saturate at some multiple $\mu_{\pm}$ of the critical \cite{Goldreich1969} flux.
	
	An important characteristic of the wind is the magnetization:
	\begin{equation}
	\sigma_0\equiv \frac{\Phi^2 \Omega^2}{\dot{M} c^3}   \approx\frac{3\dot{E}}{2\dot{M}c^2}
	\end{equation}
	where $\Phi$ is the magnetic flux of open field lines divided by $4\pi$, $\Omega$ is the spin frequency of the magnetar, and the last equality is satisfied when $\sigma_0\gg 1$ (or equivalently $\dot{E}\approx \dot{E}_{\rm mag}$). When this condition is met, $\sigma_0$ is linearly related to the energy per baryon, $\eta\equiv \frac{\dot{E}}{\dot{M}c^2}$.  The latter also equals the maximum asymptotic bulk Lorentz factor attained by the jet if the magnetic energy is fully converted into kinetic energy.
	
	The procedure to calculate the evolution of $\dot{E}(t),\sigma_0(t),\Omega(t)$ are described in \cite{Metzger+11}, given the time evolution of the neutrino luminosity\footnote{This is a weighted neutrino luminosity, accounting for both electron neutrinos and anti-neutrinos.}, $L_{\nu}(t)$, the mean neutrino energy, $\epsilon_{\nu}(t)$ and the radius, $R_{\rm NS}(t)$, of the proto-NS, as calculated by \cite{Pons+99}.  The results depend on the strength of the dipole magnetic field at the NS surface, $B_{\rm dip,0}$, the initial proto-NS spin period, $P_0$ (defined as the minimum spin period the proto-NS would achieve were it to contract to its final radius at fixed angular momentum; \citealt{Metzger+11}), the location of the light-cylinder $R_L= c/\Omega = c P(t)/2\pi $ relative to the radial extent of the closed magnetosphere, the magnetic obliquity, $\chi$, the NS mass, $M_{\rm NS}$, and the pair multiplicity of positrons/electrons produced by magnetospheric acceleration, $\mu_{\pm}$.  As fiducial values, we adopt $P_0=1.5$ ms, $\chi=\pi/2$, $M_{\rm NS}=1.4M_{\odot}$, and $\mu_{\pm}=10^6$.  
	
\begin{figure*}
		\centering
		\includegraphics[scale=0.35]{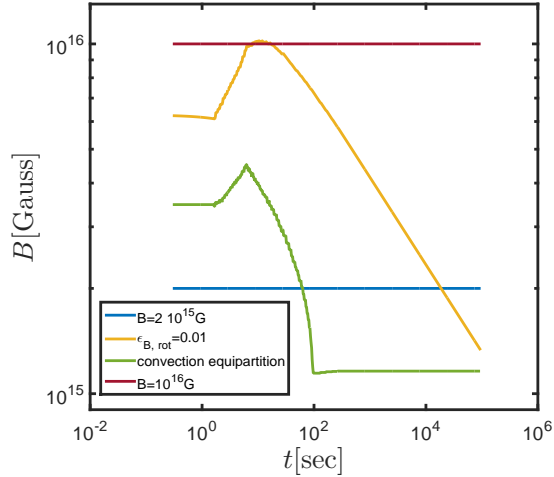}
		\caption
		{\small Four models for the time evolution of the NS surface dipole magnetic field considered in this paper: (i) temporally constant field $B=2\times 10^{15}$G; (ii) dipole magnetic energy is a fraction $\epsilon_{B,\rm rot}=0.01$ of the instantaneous proto-NS rotational energy; (iii) dipole magnetic energy density is in equipartition with the energy density of convective motions within the proto-NS; (iv) a temporally constant field $B=10^{16}$G.}
		\label{fig:differentB}
	\end{figure*}

	We consider four separate models for the time evolution of the surface dipole magnetic field strength, as summarized in Figure \ref{fig:differentB}:
	\begin{enumerate}
		\item Temporally constant value of $B_{\rm dip,0}=2\times 10^{15}$ G.
		\item Dipole magnetic energy at all times a constant fraction $\epsilon_{B,\rm rot} = 0.01$ of the NS rotational energy: \begin{equation}
		B_{\rm dip,0}=\bigg(\frac{3\epsilon_{B,\rm rot} I_{\rm NS}\Omega^2}{R_{\rm NS}^3} \bigg)^{1/2}
		\end{equation}
		where $I_{\rm NS}$ is the moment of inertia of the NS.
		\item Energy density of the dipole magnetic field in equipartition with the convective energy density of the proto-NS (e.g.~\citealt{Sukhbold&Thompson17}).  The latter is estimated by the requirement to carry outwards the neutrino luminosity according to
		\begin{equation}
		B_{\rm dip,0}=(4\pi \rho_{\rm NS} v_{\rm conv}^2)^{1/2}
		\end{equation}
		where $\rho_{\rm NS}$ is the average NS density and
		\begin{equation}
		v_{\rm conv} \approx \bigg(\frac{L_{\nu}}{4\pi R_{\rm NS}^2\rho_{\rm NS}}\bigg)^{1/3}
		\end{equation}
		is the convective velocity.  The proto-NS becomes optically thin to neutrinos at $t \approx 100$ s, after which time the neutrino luminosity abruptly drops to very low values; we assume that after this transition the magnetic field remains frozen at its value just prior to the cut-off of the neutrino emission.
		\item Temporally constant value of $B_{\rm dip,0}=10^{16}$ G.
	\end{enumerate}
	Although none of these models are likely to be correct in detail, they provide a representative range of possibilities.  Within the range of time-scales relevant for prompt GRB emission, the constant magnetic fields of $2\times 10^{15}$ and $10^{16}$ G provide approximate lower and upper limits, respectively, to the more ``physically-motivated" models (2) and (3).

	\subsection{Wind Evolution and General Properties}
	
\begin{figure*}
		\centering
		\includegraphics[scale=0.8]{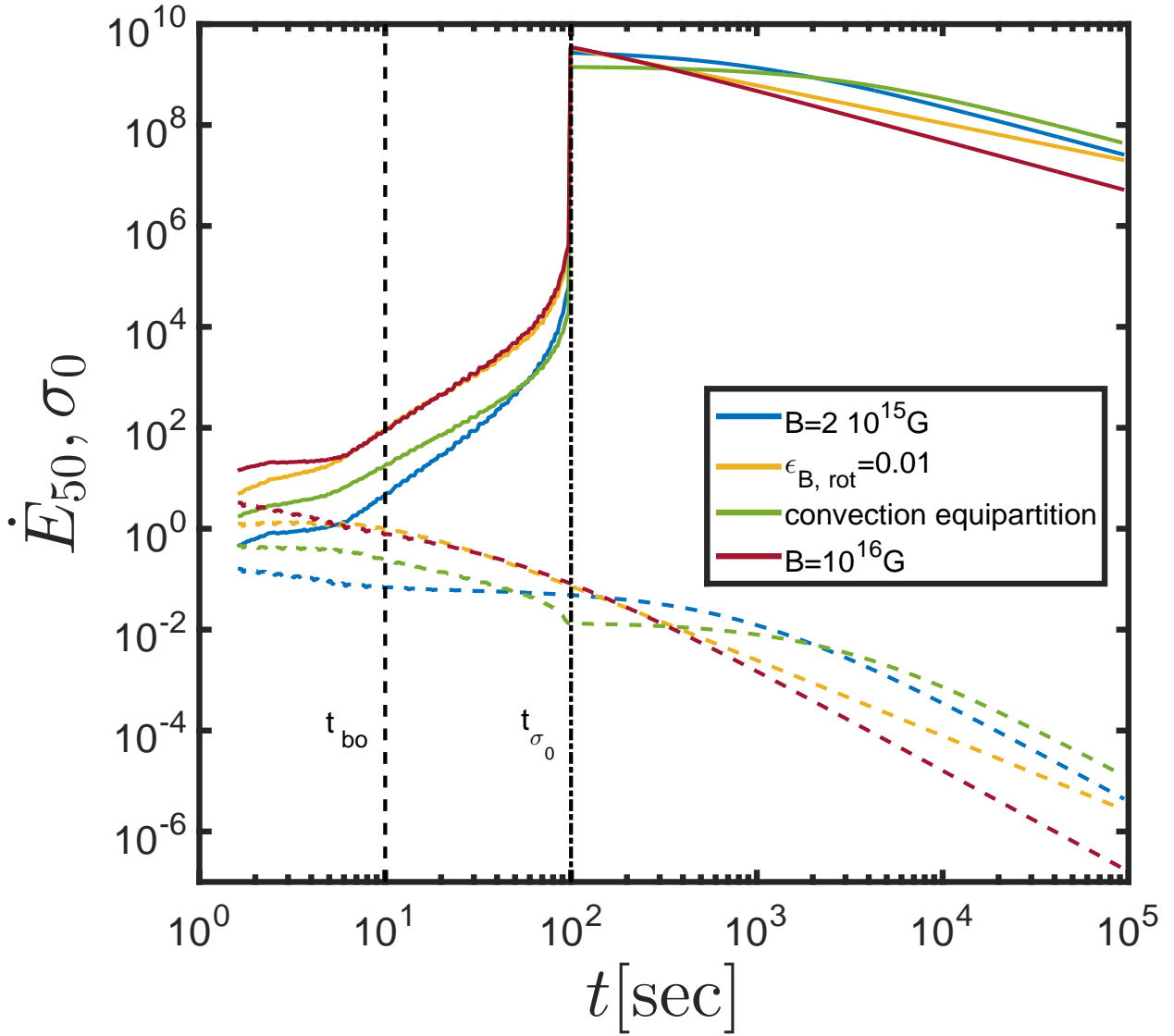}
		\caption
		{\small Evolution of the wind magnetization $\sigma_0(t)$ (top curves, solid lines) and total luminosity $\dot{E}(t)$ (bottom curves, dashed lines), shown the different magnetic field evolution models discussed in \S \ref{sec:magnetarevolve}.  The approximate time for the jet to break out of the star, $t_{\rm bo}$ and the time at which the wind undergoes the high $\sigma_0$ transition, $t_{\sigma_0}$ are denoted by dashed and dot-dashed lines, respectively.}
		\label{fig:Edotnsigma}
	\end{figure*}

	The evolution of $\dot{E}(t),\sigma_0(t)$ for the four magnetic field models are shown in Fig. \ref{fig:Edotnsigma}.  The stronger magnetic fields in cases (2) and (4) yield correspondingly larger values of $\dot{E}$, resulting in an effective spin-down time of $\sim 100$ s (see Eq. \ref{eq:tSD}).  This timescale being of the order of tens of seconds is a necessary condition in order to extract a significant amount of energy from the magnetar, between the moment the jet breaks out of the star, $t_{\rm bo} \approx 10$ s and the time of the ``high $\sigma_0$ transition" when the proto-NS becomes optically thin to neutrinos (denoted $t_{\sigma_0} \approx$ 100 s).  As described in detail later, the latter transition likely denotes the termination of the prompt emission phase because the jet becomes radiatively inefficient once $\sigma_0$ becomes very high.
	
\begin{figure*}
		\centering
		\includegraphics[scale=0.4]{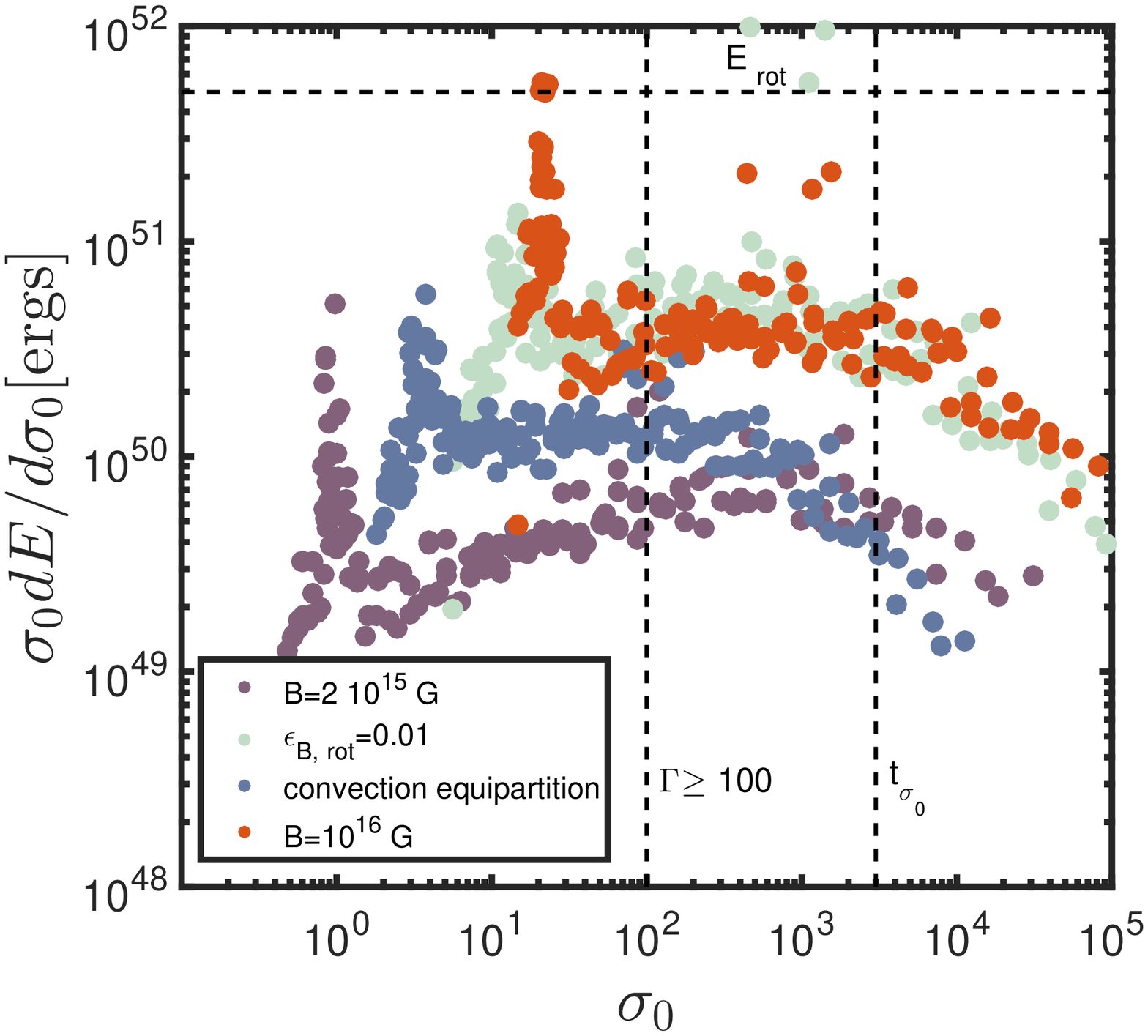}
		\includegraphics[scale=0.4]{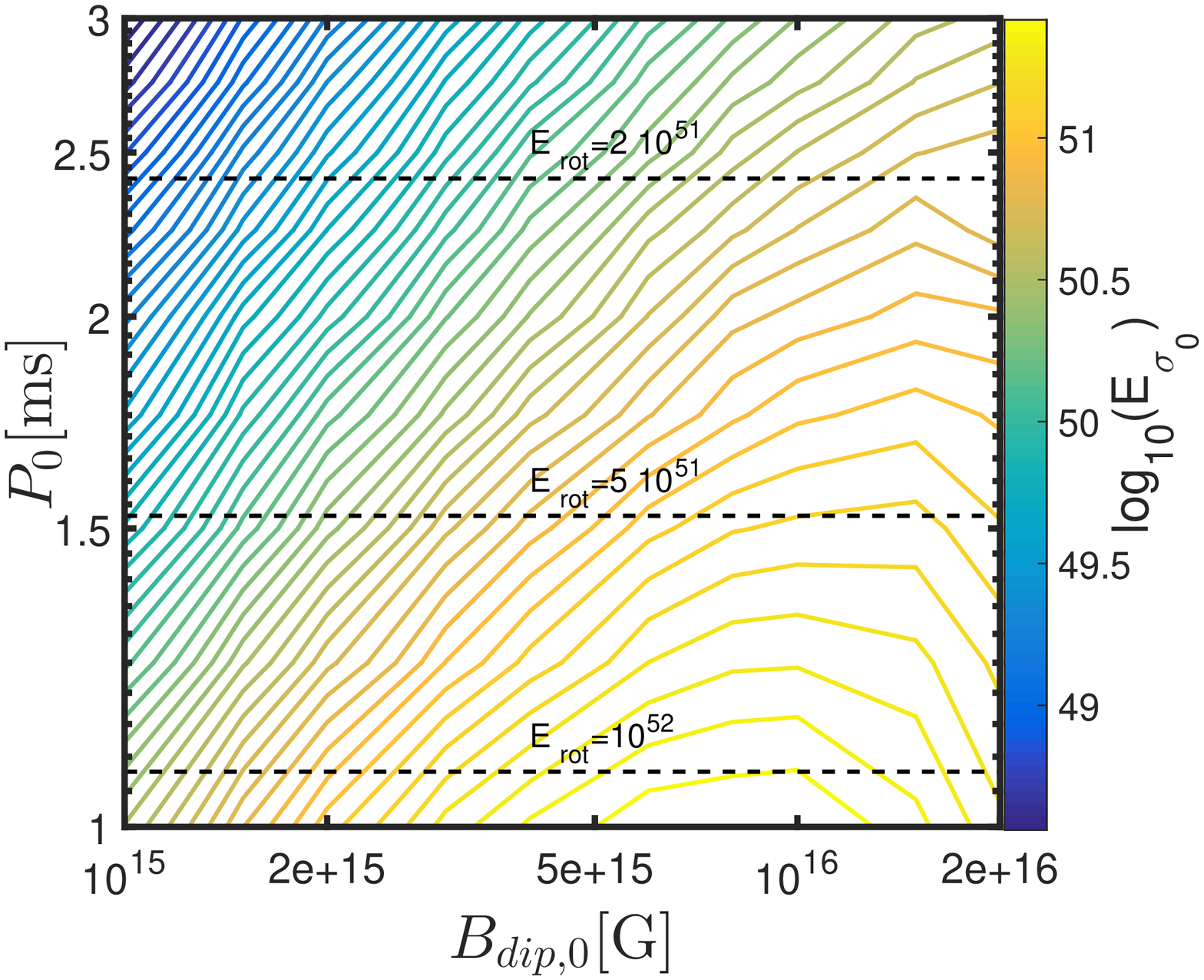}
		\caption
		{\small {\bf Left:} Distribution of magnetar rotational energy released per decade of wind magnetization, $\sigma_0 dE/d\sigma_0$, shown for the different magnetic field models introduced in \S \ref{sec:magnetarevolve} and a magnetar with $M_{\rm NS}=1.4M_{\odot}$. {\bf Right:} Total rotational energy $E_{\sigma_0}$ released by the magnetar wind in the magnetization range $\sigma_0=100-3000$, for different values of the (temporally constant) surface dipole magnetic field and initial spin period. These values represent model-independent approximate upper limits on the total available energy for powering a GRB from a magnetar central engine. Horizontal lines depict the rotational energy reservoir.  In all cases $E_{\sigma_0}\leq 0.25 E_{\rm rot}$.}
		\label{fig:dEdsigma}
	\end{figure*}

	The left panel of Fig. \ref{fig:dEdsigma} shows the distribution of the total energy released by the wind per logarithmic decade of magnetization $\sigma_0$.  Most of the magnetar rotational energy is released between $20\lesssim \sigma_0 \lesssim 3000$ for $B_{\rm dip,0}=10^{16}$G ($1\lesssim \sigma_0 \lesssim 2000$ for $B_{\rm dip,0}=2\times 10^{15}$G), with the energy released per logarithmic interval $\sigma_0$ being relatively constant across this magnetization range.
	
	The dramatic time evolution of $\sigma_0$ makes it the main parameter controlling the GRB prompt evolution within the magnetar model.  The value of $\sigma_0$ also represents the maximum possible jet Lorenz factor $\Gamma$ at a given time achieved if all magnetic energy is converted to bulk kinetic energy.  Because compactness constraints on GRB prompt emission demand $\Gamma \gtrsim 100$, then $\sigma_0 \gtrsim 100$ represents a model-independent lower limit on the jet magnetization during the prompt emission phase.  A generic upper limit on the total available energy for powering the prompt GRB is therefore obtained by calculating the rotational energy extracted between the time when $\sigma_0=100$ and ending when $\sigma_0$ shoots up rapidly at $t_{\sigma_0}$\footnote{We define this transition as $\sigma_0=3000$; however, due to the extremely rapid rise in $\sigma_0$, the result is insensitive to this choice.}, a quantity we denote as $E_{\sigma_0}$.
	
	The values of $E_{\sigma_0}$ for different magnetic fields and initial spin periods are shown in the right panel of Fig. \ref{fig:dEdsigma}. At low $B_{\rm dip,0}$ and large $P_0$ the magnetar has not spun down significantly by $t_{\sigma_0}$ and thus $E_{\sigma_0}$ is much smaller than the total initial rotational energy $E_{\rm rot}$, i.e.~ $E_{\sigma_0}\approx E_{\rm rot}(t_{\sigma_0}/t_{\rm SD} )$, where $t_{\rm SD}$ is the spin-down time (see Eq. \ref{eq:tSD}).  At larger $B_{\rm dip,0}$ and lower $P_0$, the value of $E_{\sigma_0}$ is still less than $E_{\rm rot}$ because it is now limited by $t(\sigma_0=100)$ since the magnetar releases a significant fraction of its initial energy before this time.  In all considered cases, $E_{\sigma_0}\lesssim 0.25 E_{\rm rot}$, with a maximum value of $E_{\sigma_0}\approx 3\times 10^{51}$erg obtained for $P_0=1$ms and $B_{\rm dip,0}=2\times 10^{16}$G.  This represents an approximate upper limit on the radiative efficiency of the magnetar model, independent of the emission mechanism.
	
	These estimates come with several caveats.  One includes the fact that neutrino-driven mass loss from the proto-NS simply sets a minimum baryon loading of the jet; other processes, such as entrainment from jet walls, may well come to dominate at late times once $\sigma_0$ becomes sufficiently large.  The true evolution of $\sigma_0(t)$ could therefore be shallower than we have estimated based on the proto-magnetar wind evolution alone.  Our results for $\sigma_0(t)$ are furthermore sensitive to the details of the proto-NS cooling evolution, which to date are based on one-dimensional stellar evolution models which neglect potentially important effects of rapid rotation and strong magnetic fields \citep{Pons+99,Roberts+12}. A shallower evolution of $\sigma_0$ would mainly affect the resulting bursts' properties at late times. If this affect is sufficiently strong, it could increase the timescale of GRBs in this model as well as the total energy than can be extracted from the magnetar.
	
	\section{Prompt Emission Models}
	\label{sec:GRB}
	Predicting the GRB emission properties from magnetar central engines requires knowledge of where the dissipation of ordered magnetic or kinetic energy takes place within the jet.
	In the following sections we consider three possibilities for the energy dissipation which cover, respectively, the possibilities of (1) dissipation that peaks close to the central engine (``fireball", \S \ref{sec:Fireball}); (2) dissipation of kinetic at large distances via internal shocks (\S \ref{sec:intShock}); and (3) gradual magnetic dissipation occurring across a range of radii throughout the jet (\S \ref{sec:magdiss}).

	\subsection{Pure Fireball}
	\label{sec:Fireball}
	We consider first a jet for which the dynamics are that of a pure hot fireball \citep{Goodman1986,Paczynski1986}, the energy of which is supplied by the proto-magnetar.  This case corresponds to the limit of catastrophic reconnection in the flow on small scales near the proto-NS or the light cylinder, i.e. prior to significant acceleration of the jet out of the star.  
	
	Near the base of the jet at radius $R_{0}$, the isotropic equivalent luminosity is $L_{\rm iso}\equiv \dot{E} f_b^{-1}$ (where $f_b$ is the solid angle covered by the jet over $4\pi$) and the outflow temperature is given by
	\begin{equation}
	\label{eq:T0}
	k_B T_0\approx k_B \bigg( \frac{L_{\rm iso}}{4\pi R_0^2 g_0 \sigma_B} \bigg)^{1/4}=1.3\mbox{MeV} L_{\rm iso,52}^{1/4}R_{0,7}^{-1/2}
	\end{equation}
	where $k_B,\sigma_B$ are the Boltzmann and Stefan-Boltzmann constants, and $g_0\approx 2.75$ is half the degrees of freedom for a photon, electron and positron plasma (e.g. \citealt{KumarZhang2015}).  Here and throughout, subscripts denote normalization to cgs units, e.g. $R_{0,7} \equiv R_{0}/10^{7}$ cm.
	
	Denote the radius of a given propagating shell by $r$, and its thickness by $\delta r$. The latter is approximately constant as both edges of the shell are moving nearly the same velocity close to the speed of light.  As the fireball expands adiabatically, its luminosity in the observer frame obeys $L_{\rm iso}\propto r^2 T'(r)^4 \Gamma (r)^2$ (where primes are used to denote quantities in a frame that is co-moving with the jet), while its entropy (which is frame independent), $S\propto r^2 \delta r' T'(r)^3$, remains constant \citep{Goodman1986,Paczynski1986}.  Solving these equations gives $\Gamma \propto r$ and $T' \propto r^{-2/3} \Gamma^{-1/3}\propto r^{-1}$.
	
	This solution holds until either (1) $\Gamma(r)$ saturates at a value $\eta$ equal to the total energy per baryon at the base of the jet at the radius $R_{s}$ (after which point no more energy is available to further accelerate the jet) or (2) the jet reaches its photosphere and the energy is lost to radiation, at which point $\Gamma$ saturates at a value $\ll \eta$. Comparing the photosphere radius
	\begin{equation}
	\label{eq:photo}
	R_{\rm ph}=\frac{L_{\rm iso} \sigma_T}{8\pi m_p c^3 \eta \Gamma^2}\approx 5.5\times 10^{12}\mbox{cm} L_{\rm iso,52}\eta_2^{-1} \Gamma_2^{-2}.
	\end{equation}
	to the saturation radius $R_s=R_0 \eta$, and defining as $\eta_*$ the value of $\eta$ for which the two radii are equal ($R_{\rm ph}(\eta_*)=R_s(\eta_*)$), delineates these two possibilities:
	\begin{enumerate}
		\item $R_s>R_{\rm ph} \rightarrow \eta> \eta_* \sim 8.5 \times 10^2 L_{\rm iso,52}^{1/4} R_{0,7}^{-1/4}$. In this case the jet will radiate the majority of its energy at the photosphere.  Since the photosphere lies here below the saturation radius, the observed luminosity and temperature will be approximately equal to those near the base of the jet.
		\item $R_s<R_{\rm ph} \rightarrow \eta< \eta_* \sim 8.5 \times 10^2 L_{\rm iso,52}^{1/4} R_{0,7}^{-1/4}$.  In this case the temperature and luminosity remain constant up until $R_s$.  Above this radius $\Gamma$ saturates to its terminal value and the jet thermal energy begins to decline because of its expansion, such that $T \propto r^{-2/3}, L_{\rm iso} \propto r^{-2/3}$.  The residual thermal energy is radiated only once the jet reaches its photosphere, resulting in a lower radiative efficiency than case 1.
	\end{enumerate}
	
	\subsection{Internal Shocks}
	\label{sec:intShock}
	Here we assume that the jet accelerates efficiently, such that it directly converts a large fraction of its Poynting flux into kinetic energy without dissipating energy into particle heating at small radii.  In this case the engine power may be dissipated at larger radii when different velocity components (``shells'') within the jet collide with each other and create shocks.  
	
	For purposes of isolating potential shock-powered emission, we assume that the jet converts 100\% of its energy to kinetic form (with zero radiative losses).  The jet kinetic luminosity and bulk Lorentz factor are thus given by $L_j(t)=\dot{E}(t)$ and  $\Gamma_j(t)=\eta(t) = \sigma_0(t)$, respectively.  Because $\sigma_0  = \eta$ rises with time as the magnetar cools, the jet at later times has larger $\Gamma_j$ and thus a collision is expected when the fast material catches up with previously ejected slower material. This collision becomes stronger with time as the relative Lorentz factor between the new material and the previously-accumulated 'bulk' shell increases.  The shocks accelerate electrons in the shell to relativistic energies and amplify the  magnetic field, radiating the electron energy as synchrotron emission.
	
	We apply the formulation of \cite{Metzger+11} to calculate the radiative luminosity and synchrotron peak energy as a function of time. As shown in that work, at late times, one can approximate these two quantities as
	\begin{equation}
	L_{\rm rad,iso}(t)\approx \frac{\epsilon_e}{2}L_j(t) f_b^{-1}
	\end{equation}
	\begin{equation}
	E_{\rm peak}(t)=4.7\mbox{MeV}\frac{\epsilon_e^2 \epsilon_B^{1/2}}{\xi^2}\!\bigg(\frac{L_{j,\rm iso}}{10^{51}\mbox{erg s}^{-1}}\!\bigg)^{1/2}\!\bigg(\frac{10 s}{t}\!\bigg)\!\Gamma_j^2\Gamma_s^{-4}
	\end{equation}
	where $\epsilon_e, \epsilon_B$ are the fraction of shock dissipated energy placed into electrons and magnetic fields, respectively ($\epsilon_B$ here should not be confused with $\epsilon_{B,\rm rot}$ defined earlier when discussing the NS dipole magnetic field evolution), $\xi$ is the fraction of electrons accelerated, $L_{j, \rm iso}\equiv L_j f_b^{-1}$ and $\Gamma_s$ is the Lorentz factor of the bulk shell.
	
	We consider also a slight variation on the above model, in which, on top of the gradually rising value of $\eta(t)$, we introduce a random short time variability in $\log(\eta)$ with $\Delta t=1$ s and $\Delta \eta/\eta=2$, motivated by observations of GRB variability.  These variations result in internal shocks at a distance $\sim \mbox{min}(\eta(t)^2 c \Delta t,2\Gamma_s^2 c t)$, depending on whether matter first collides with itself or with the bulk shell.  The shock distance is generally dominated by the rapid variability for the first 10$-$30 s, after which point $\eta(t) \gg \Gamma_s$ and the shock distance is set by the slow secular growth of $\eta(t)$.
	
	\subsection{Gradual Magnetic Dissipation}
	\label{sec:magdiss}
	
	In Poynting-flux dominated jets, the GRB prompt emission can be efficiently powered by magnetic reconnection.  This would be facilitated if the magnetic field contains polarity reversals in the jet, as might be expected for a magnetar with its dipole axis misaligned with respect to the rotation axis; this introduces an `AC' component to the Poynting flux which results in the dissipation of magnetic energy occurring gradually over a large range of radii \citep{Drenkhahn2002}. About half of the dissipated energy serves to directly accelerate the jet while the other half is injected internally in the flow, and can be radiated away. The dissipation process stops at the `saturation radius' by which most of the magnetic energy has been released. This results in a combination of a photopsheric emission component, released as the jet becomes transparent and a non-thermal synchrotron component released between the photosphere and the saturation radius.
	
	This gradual dissipation model has been applied to describe the GRB prompt emission in several works including \cite{GianniosSpruit2005,GianniosSpruit2007,Giannios2008} and most recently in \cite{BG2017}. As shown in \cite{BG2017} the spectra is described given the jet luminosity $L$ (for ease of comparison with the other models described above, here we will consider the isotropic equivalent injected energy rate $\dot{E}_{\rm iso}=L f_b^{-1}$ rather than the rate per steradian, as in \citealt{BG2017}), the ``wavelength" of structured magnetic field in the jet divided by the reconnection rate, $\lambda/\epsilon$, the initial energy per Baryon, $\eta$, the fraction of electrons accelerated $\xi$ and the fraction of dissipated energy that goes into these electrons, $\epsilon_e$. The parameters $\dot{E}_{\rm iso},\eta,\lambda$ describe global jet properties and evolve with time as shown in \S \ref{sec:magnetarevolve} whereas the parameters $\epsilon, \xi,\epsilon_e$ are determined by the magnetic reconnection microphysics. In a Poynting-flux dominated jet, particle-in-cell simulations indicate that these parameters are expected to be confined within a relatively narrow range, of order unity, and are unlikely to evolve significantly over time (see, e.g., \citealt{Sironi+14}; \citealt{Sironi+16}).

	The main parameter controlling the spectrum and temporal evolution of the prompt emission is  the baryon loading, $\eta$ (or equivalently, $\sigma_0$, see \citealt{BG2017}). For $\eta \lesssim 175 (\lambda/\epsilon)_8^{-1/5}\dot{E}_{\rm iso,52}^{1/5}$, where subscript 8 denotes normalization to $10^{8}$ cm (a typical value for the light cylinder size), the saturation radius lies below the photosphere, and the emission is mostly thermalized. As $\eta$ increases the non-thermal component becomes more prominent. For $\eta \gtrsim 1100 \dot{E}_{52}^{1/5} (\epsilon_e/\xi)^{1/5}(\lambda/\epsilon)_8^{-1/5}$ the accelerated electrons enter the slow-cooling regime before the saturation radius, resulting in a significant reduction of the radiative efficiency of the jet.

	\begin{figure*}
		\centering
		\includegraphics[scale=0.35]{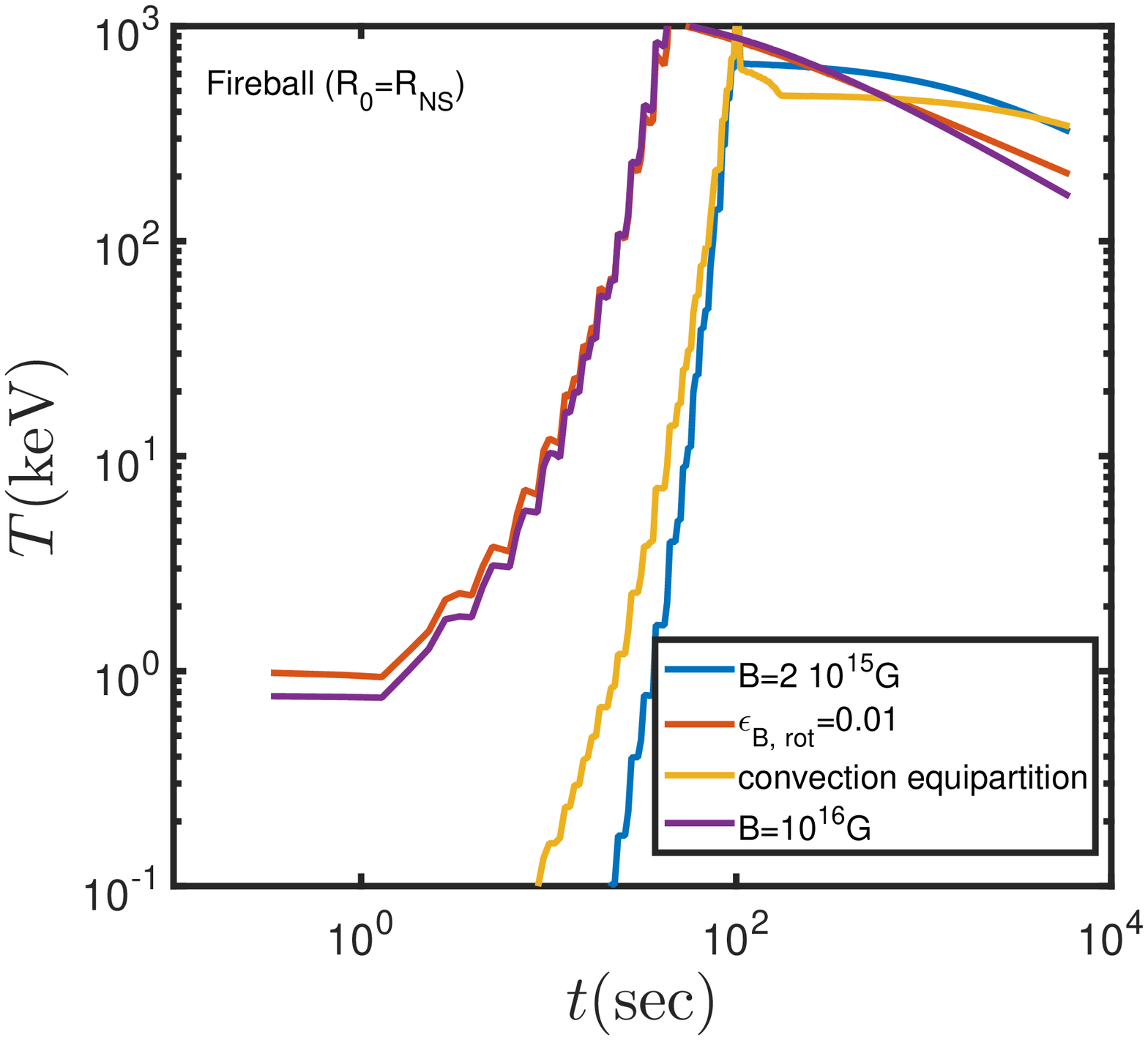}
		\includegraphics[scale=0.35]{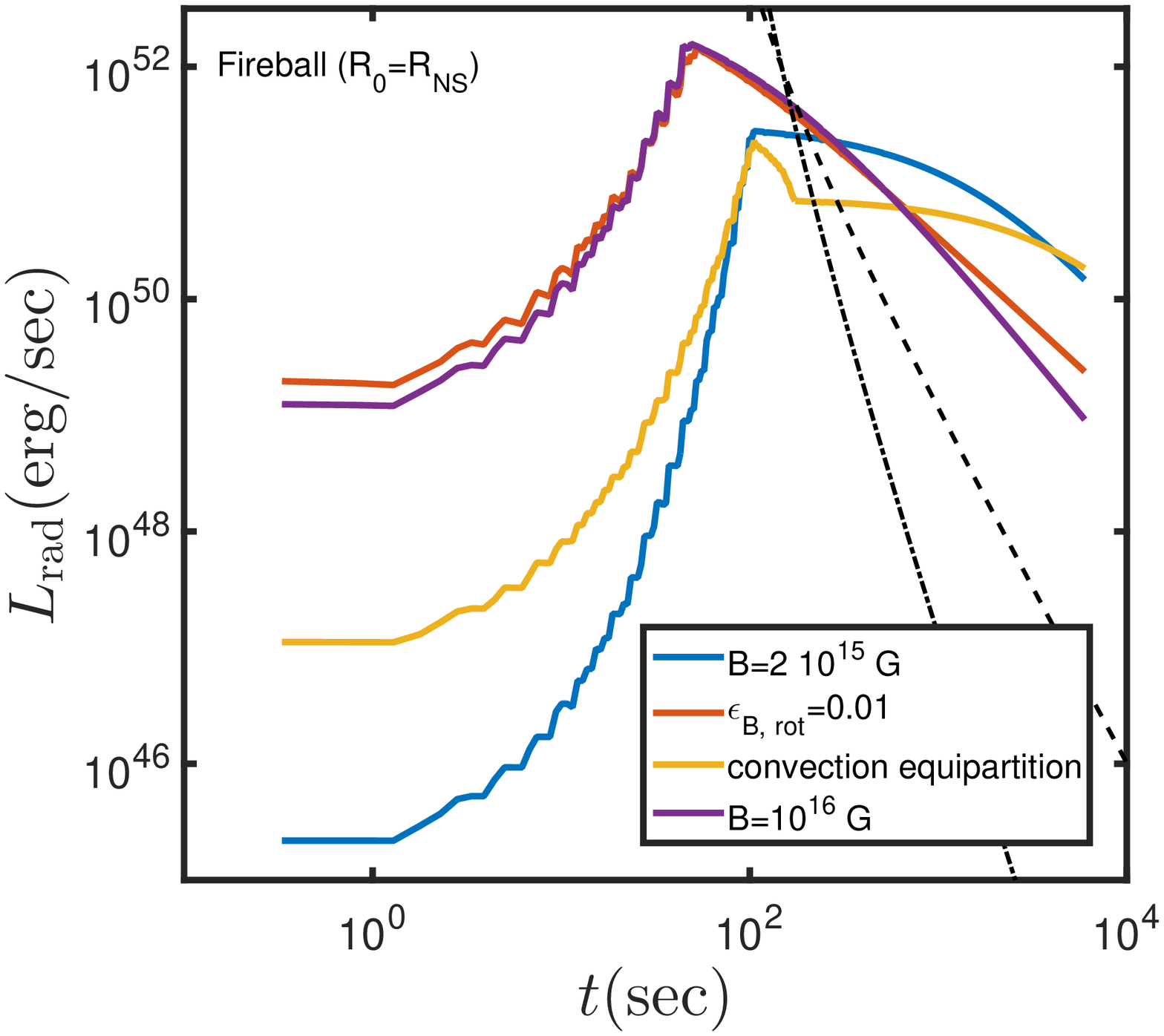}
		\caption
		{\small Evolution of observed temperature (left) and isotropic equivalent luminosity (right) in the fireball model, for different assumptions about the magnetic field evolution discussed in $\S\ref{sec:magnetarevolve}$ (Fig.~\ref{fig:differentB}). Results are shown assuming the base of the jet is located at the NS surface, and a source redshift $z=1$. Time is measured in the observer frame starting from the jet breakout time $t_{\rm bo}=10$ s. In all cases $R_s<R_{\rm ph}$ up to a few tens of seconds (during which both luminosity and temperature evolve rapidly), after which $R_s>R_{\rm ph}$ and the luminosity and temperature begin to track the same properties at the base of the jet. On the RHS, dashed and dot-dashed lines mark $(t-t_0)^{-3}$ and $(t-t_0)^{-5}$ decay rates respectively (typical of those observed during the early GRB steep decay phase), where $t_0$ is the observational trigger of the burst in this model, defined as the first time at which the luminosity reaches 0.1 of its maximum value. The predicted light curve decay is clearly too shallow compared with the observed steep-decay phase.}
		
		\label{fig:fireball}
	\end{figure*}
	
	\begin{figure*}
		\centering
		\includegraphics[scale=0.35]{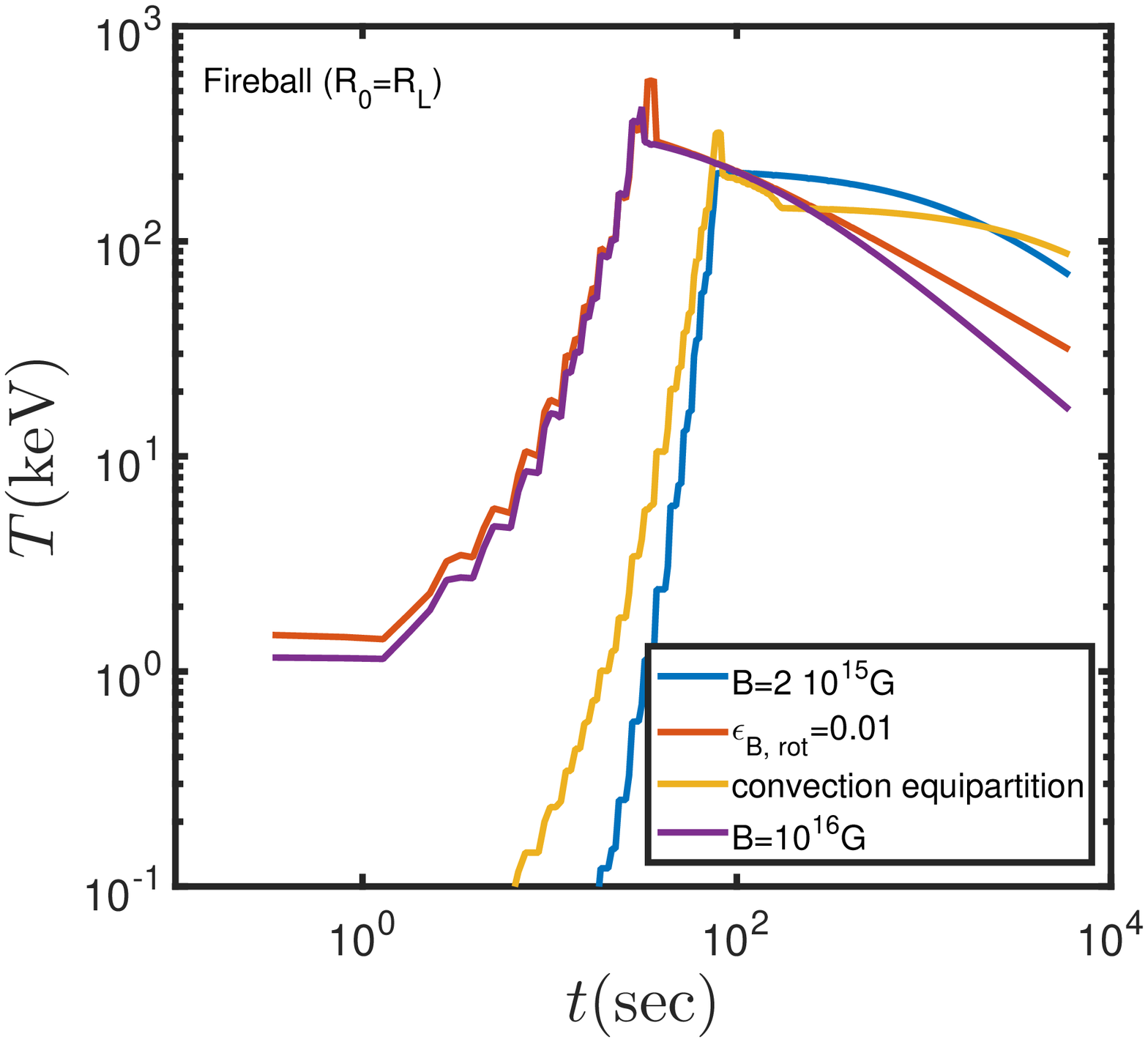}
		\includegraphics[scale=0.35]{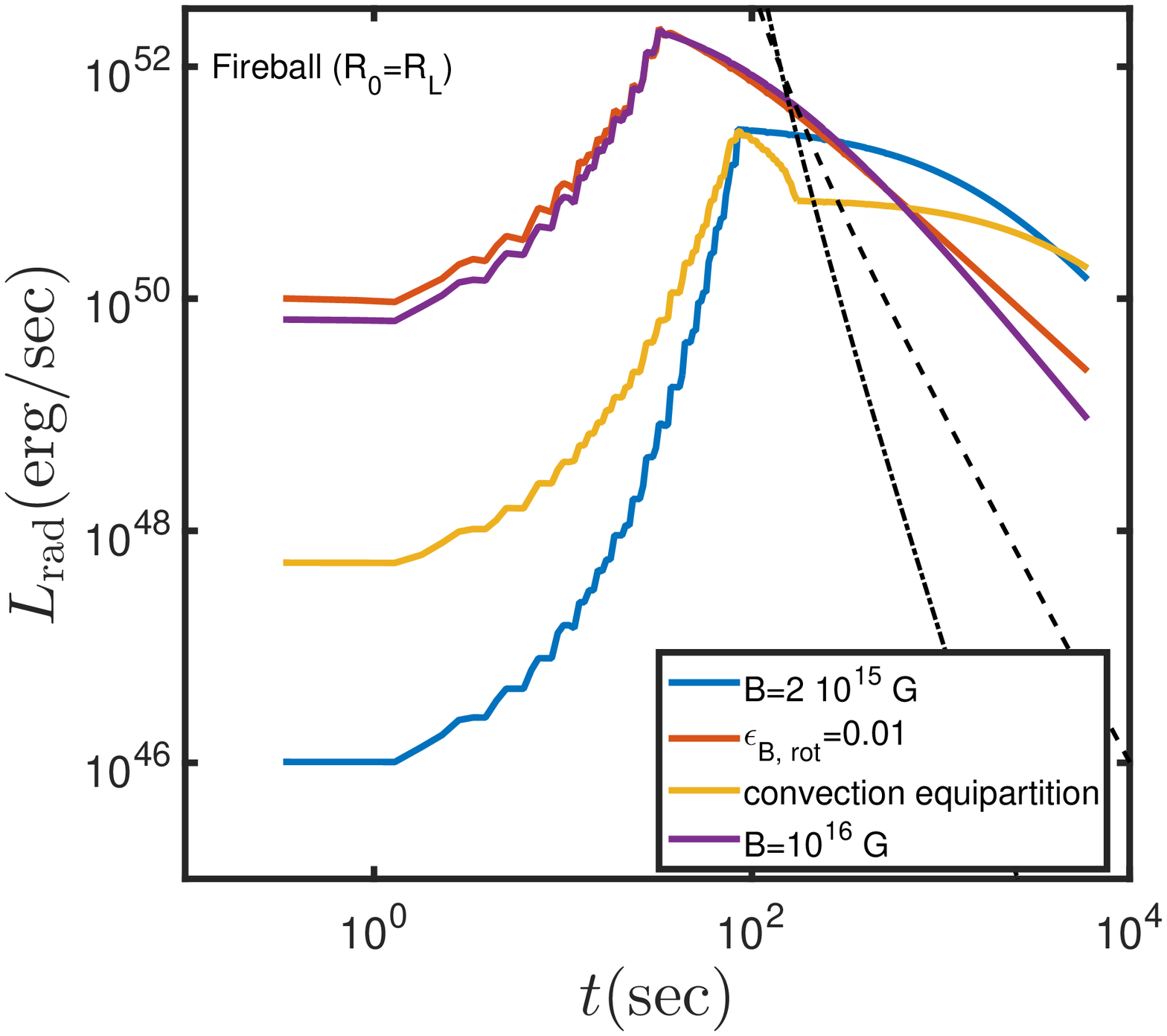}
		\caption
		{\small Same as Fig. \ref{fig:fireball}, but instead assuming the base of the jet is located at the light cylinder.}
		\label{fig:fireball2}
	\end{figure*}
	
	\begin{figure*}
		\centering
		\includegraphics[scale=0.35]{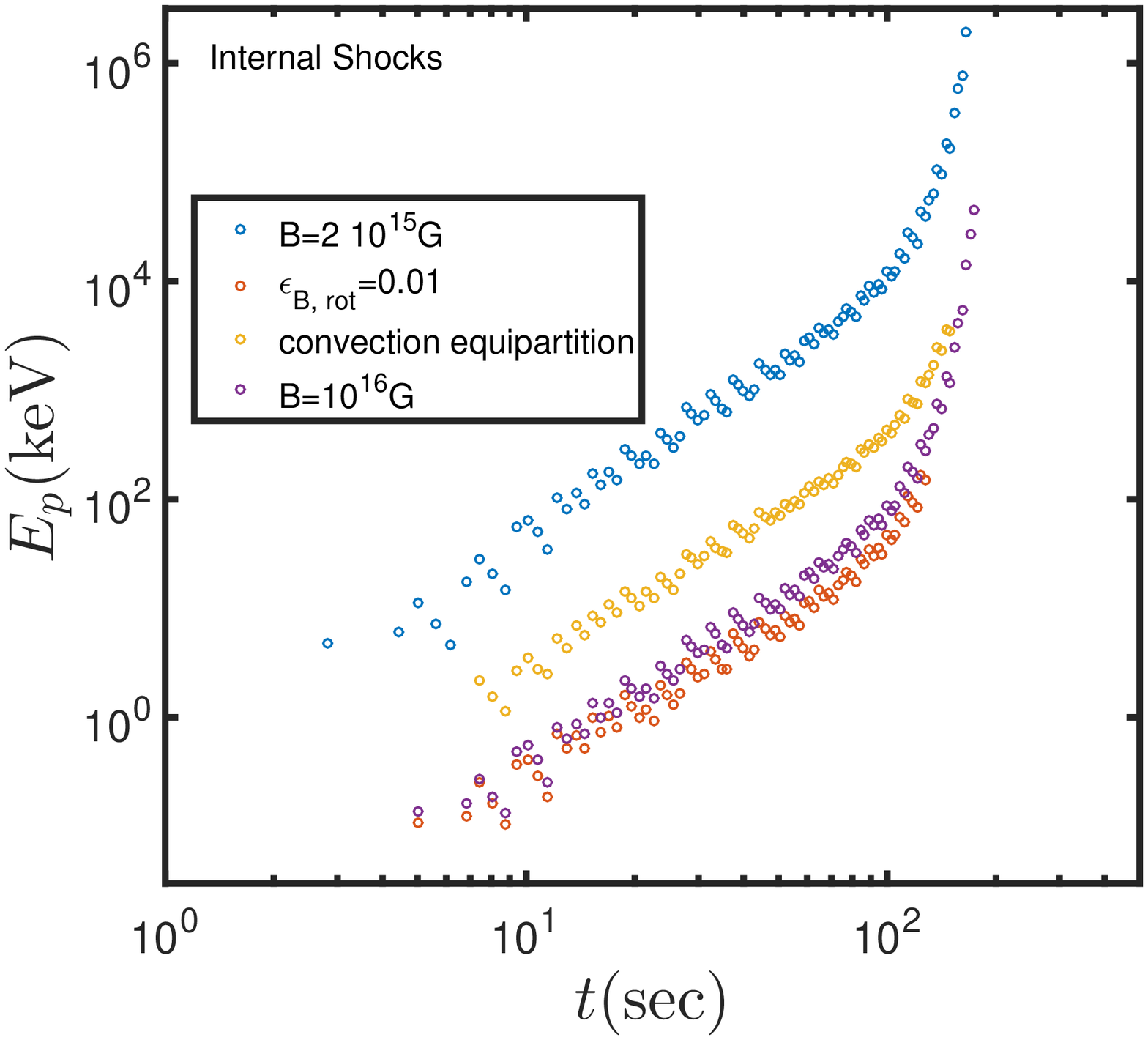}
		\includegraphics[scale=0.35]{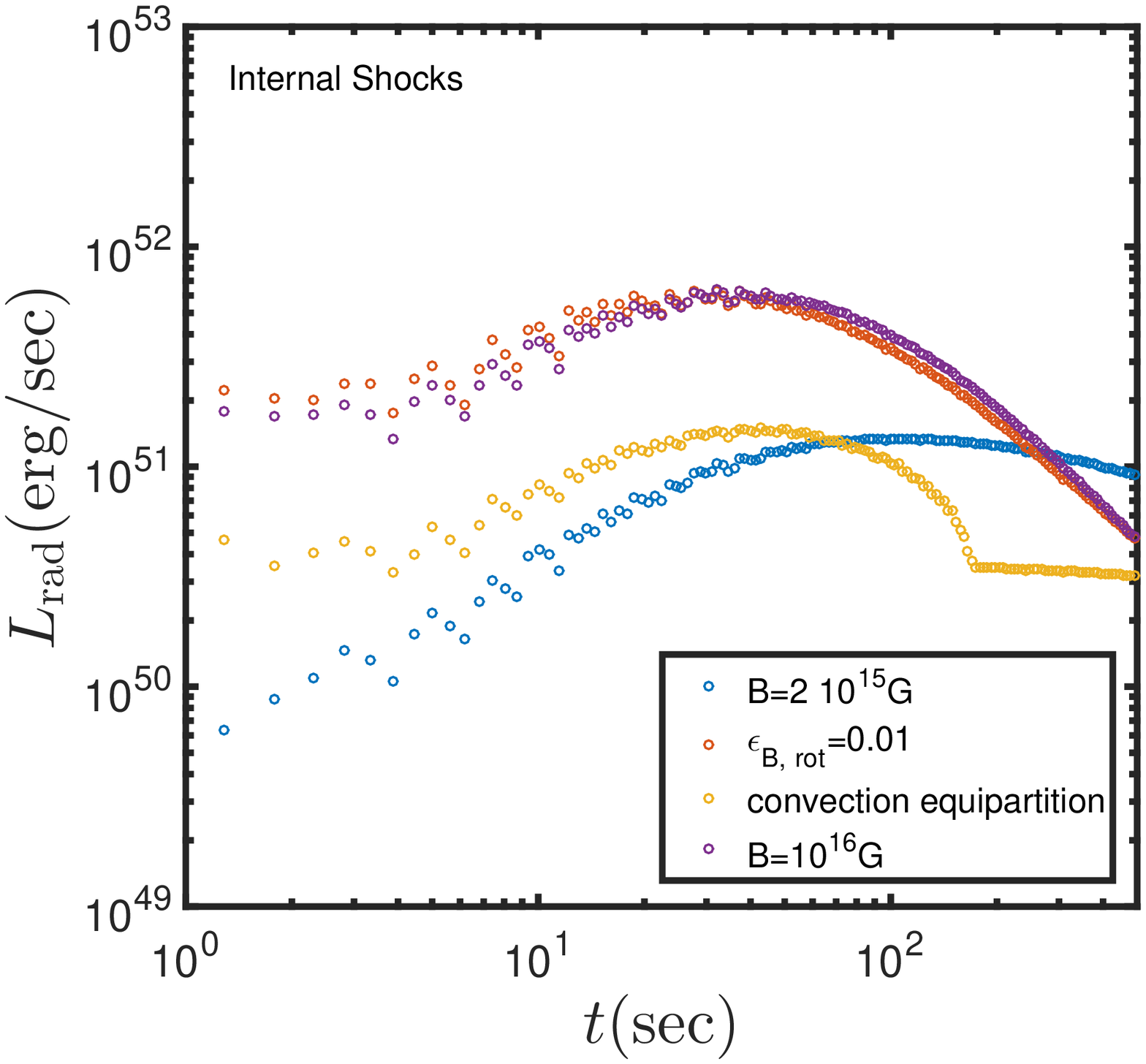}
		\includegraphics[scale=0.35]{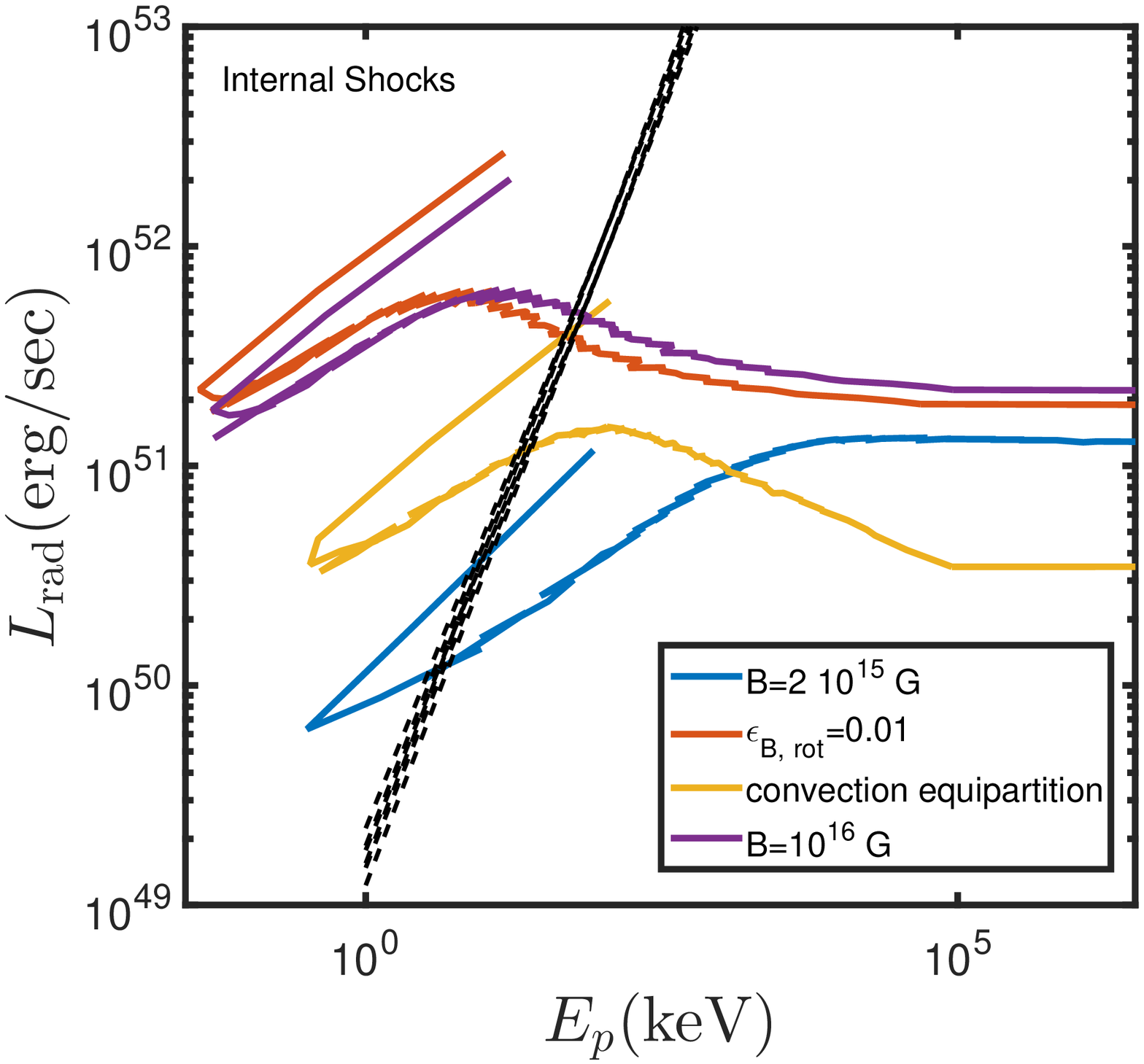}	
		\caption
		{\small Evolution of observed peak energy $E_{\rm peak}$ (left) and isotropic equivalent luminosity (right) in the internal shock model, for different magnetic field models. Results are shown assuming a jet breakout time $t_{\rm bo}=10$s and a redshift $z=1$. Time is measured in the observer frame starting from the breakout time. We also assume here: $\epsilon_e=1,\epsilon_B=0.01,\xi=0.1$. The bottom panel depicts the relation between $E_{\rm peak}$ and luminosity (both in the source frame). Dashed lines depict the observed relation between the luminosity and peak of the non-thermal part of the emission, when comparing different pulses of a given GRB \citep{Guiriec2015B}.}
		\label{fig:intshocks}
	\end{figure*}

	\begin{figure*}
		\centering
		\includegraphics[scale=0.35]{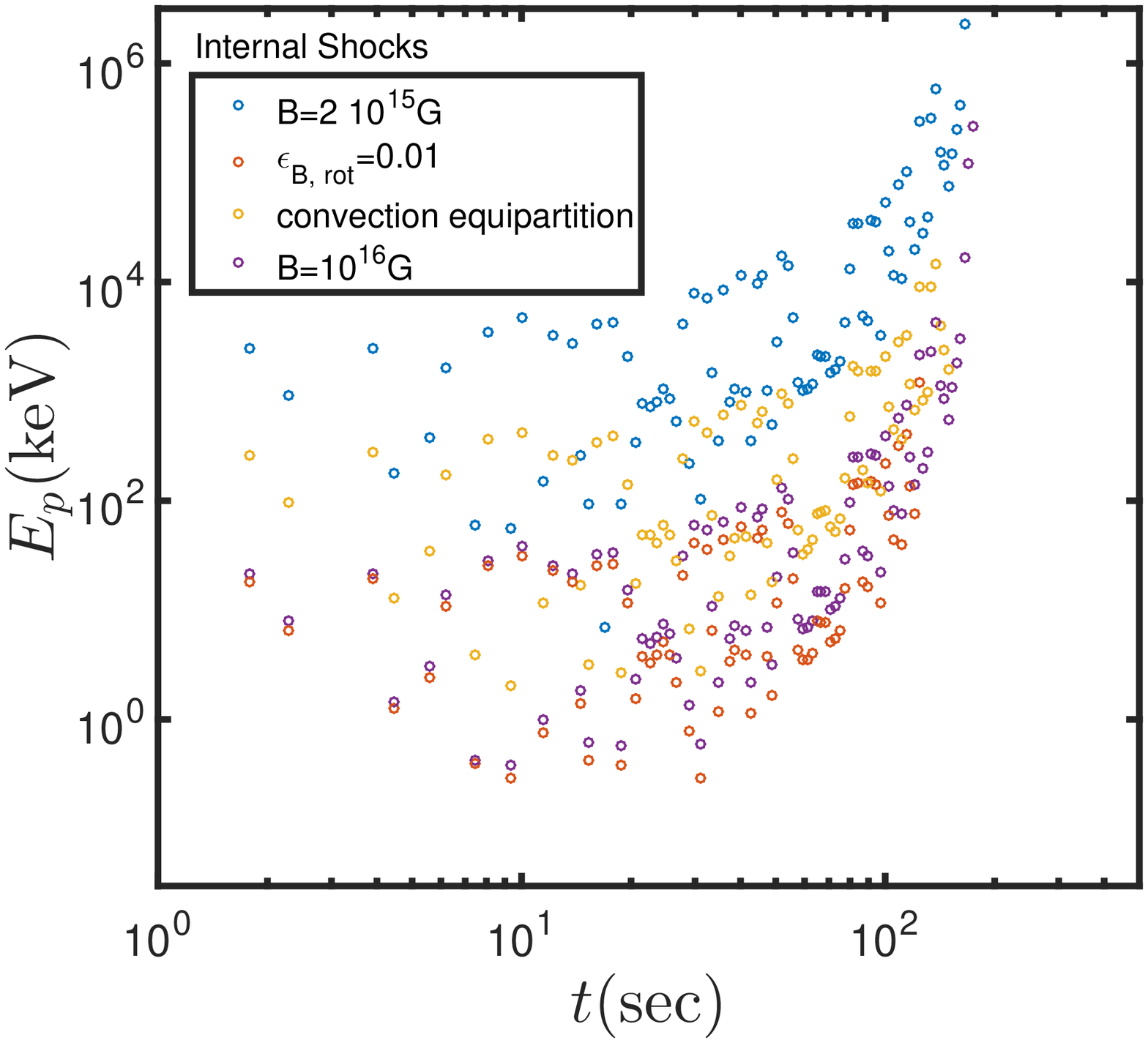}
		\includegraphics[scale=0.35]{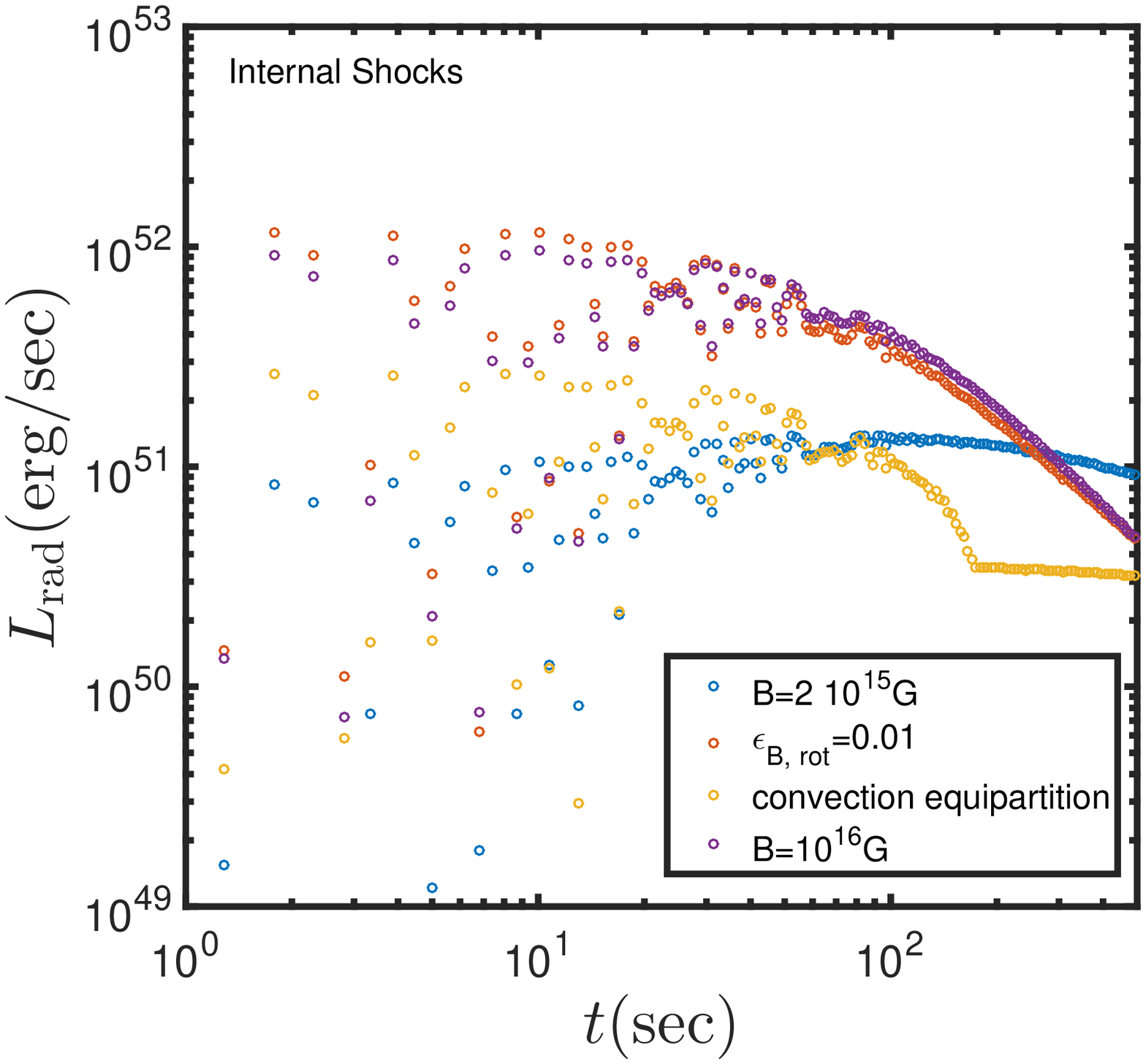}
		\includegraphics[scale=0.35]{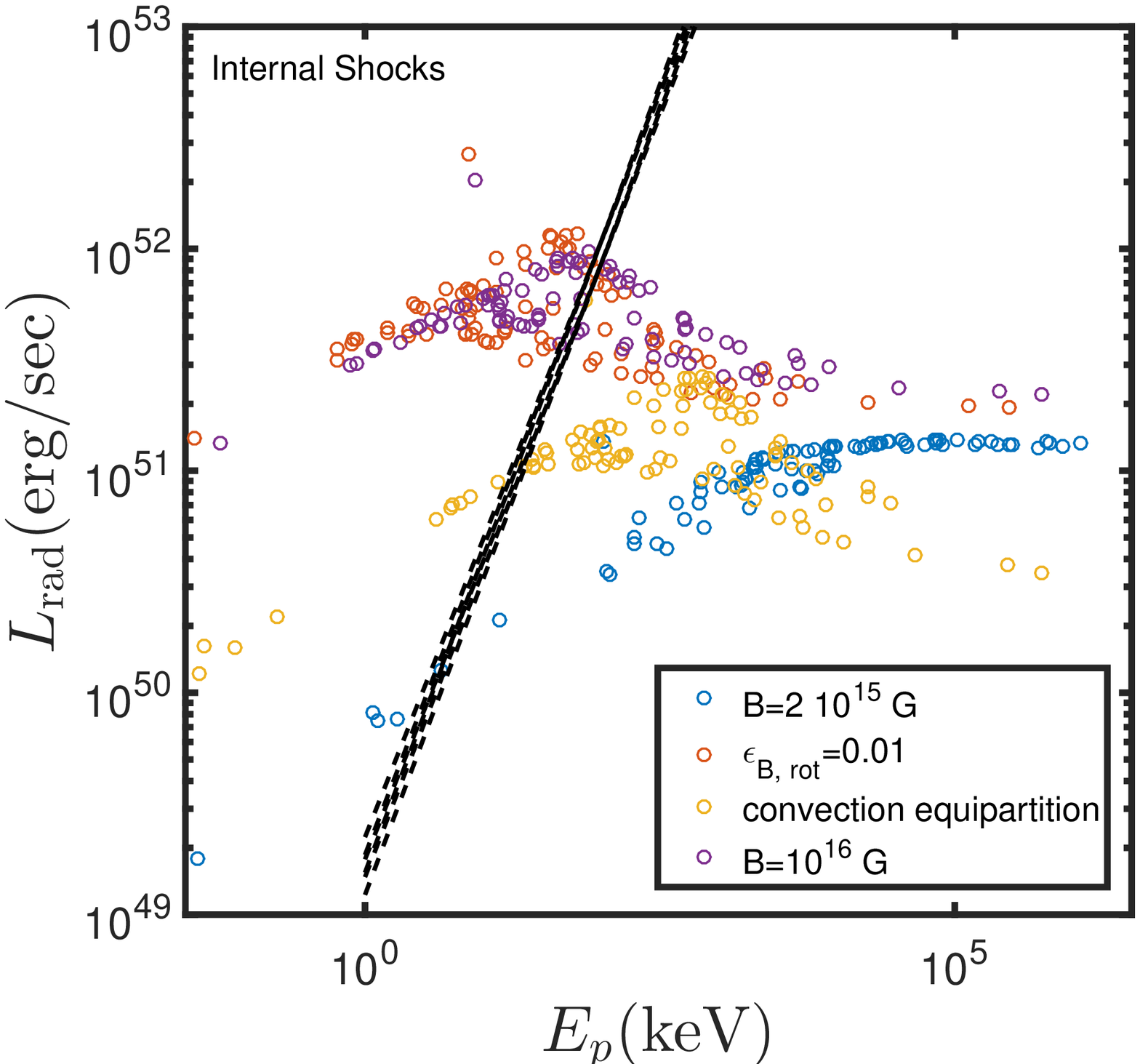}	
		\caption
		{\small Same as Fig. \ref{fig:intshocks} but for an internal shocks model where, on top of the smooth variation in $\eta(t)$, we introduce a short time variability in $\log(\eta)$ with $\Delta t=1$ s and $\Delta \eta/\eta=2$. Results for $E_p$ are only shown if $L$ at the same time is at least 0.1 of its maximum value.}
		\label{fig:intshocksvareta}
	\end{figure*}

	\begin{figure*}
		\centering
		\includegraphics[scale=0.35]{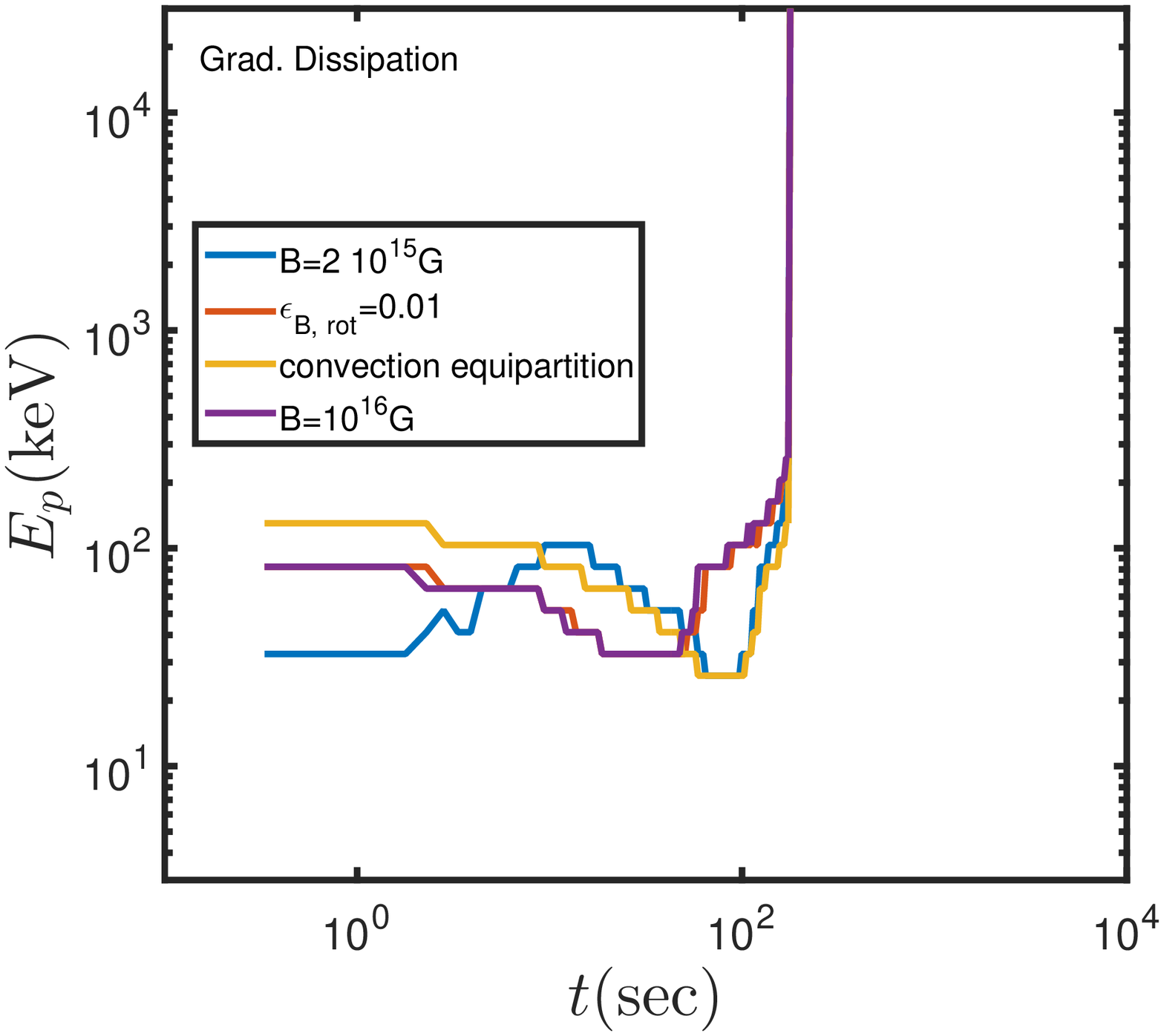}
		\includegraphics[scale=0.35]{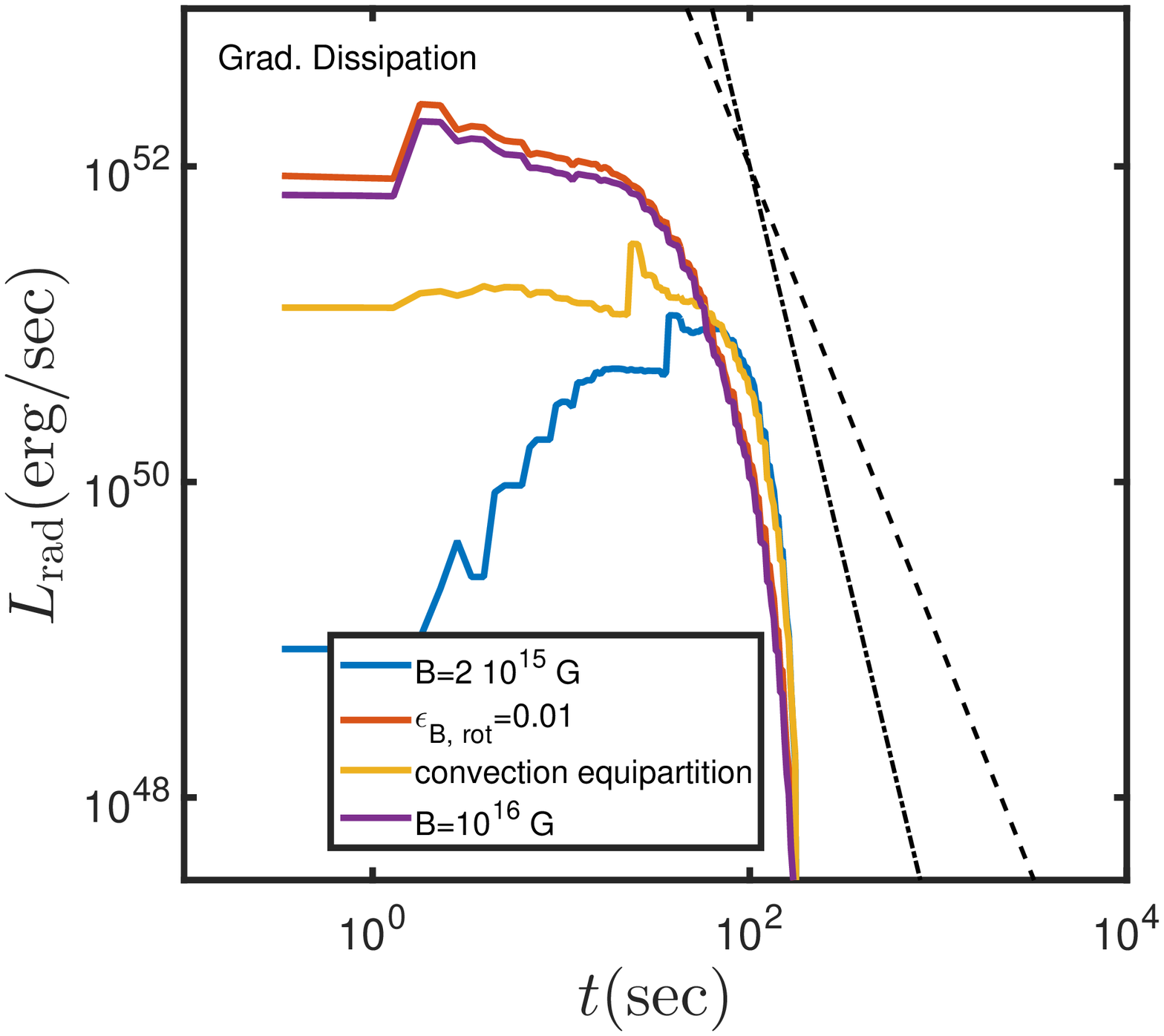}
		\includegraphics[scale=0.35]{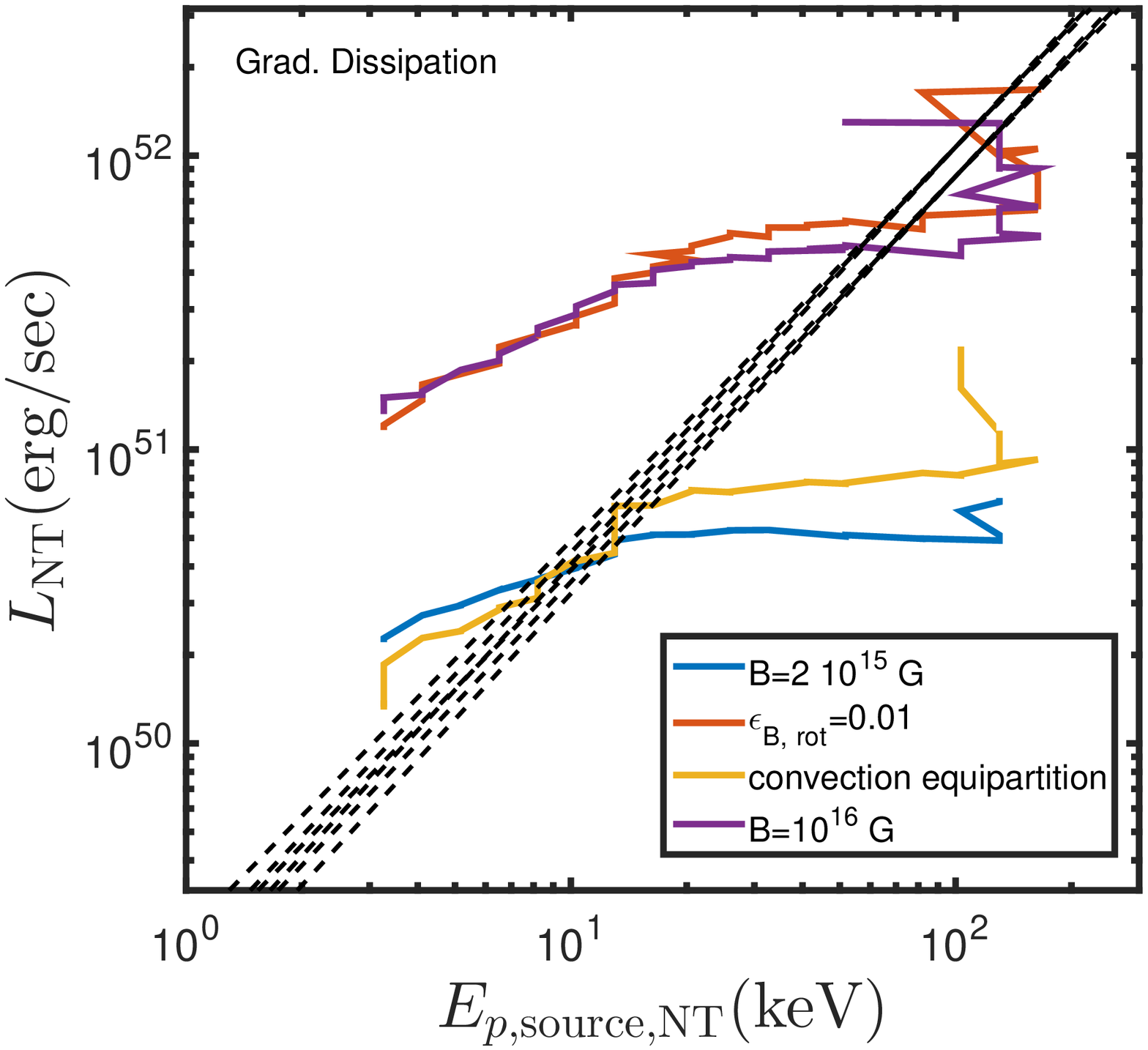}	
		\includegraphics[scale=0.35]{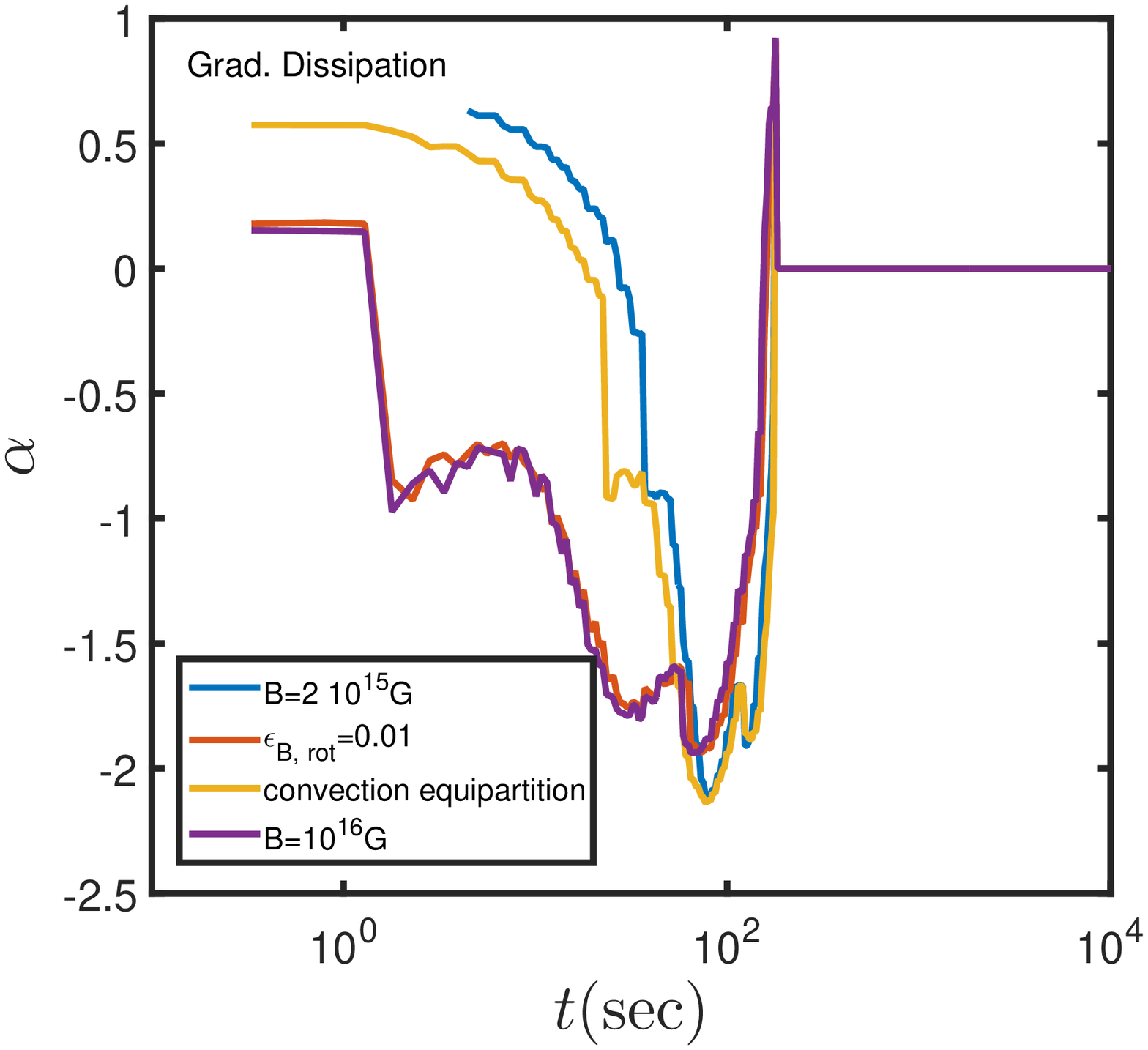}
		\caption
		{\small Evolution of observed peak energy $E_{\rm peak}$ (left) and isotropic equivalent luminosity (right) in the gradual magnetic dissipation model. Results are shown assuming a jet breakout time $t_{\rm bo}=10$s and a redshift $z=1$. Time is measured in the observer frame starting from the breakout time. We also assume here: $\epsilon_e=0.2,\xi=0.2,\epsilon=0.2$. Dashed and dot-dashed lines mark $(t-t_0)^{-3}$ and $(t-t_0)^{-5}$ decay rates respectively (typical to the observed early steep decays, as described in Fig. \ref{fig:fireball}). Bottom left: the relation between $E_{\rm peak}$ and the non-thermal luminosity (both in the source frame). Bottom right: evolution of the average photon spectral index in the 10-50 keV range.} 
		\label{fig:gradmag}
	\end{figure*}
	
	\section{Results}
	\label{sec:results}
	\subsection{Fiducial models}
	\label{sec:fiducmodel}
	Applying the evolution of $\eta(t),\dot{E}(t),\Omega(t)$ given by the magnetar model (\S \ref{sec:magnetarevolve}), we find the evolution of the observed peak energy and luminosity for the different dissipation models. We use four models for the magnetic field as specified in \S \ref{sec:magnetarevolve}. To convert from the true luminosity to the isotropic equivalent, we canonically assume a jet opening angle of $\theta_j=5^{\circ}$ \citep{Goldstein2016,Ryan2015}. We explore the dependence of the results on this assumption in \S \ref{sec:beaming}. Results are shown in Figs. \ref{fig:fireball},\ref{fig:fireball2},\ref{fig:intshocks},\ref{fig:intshocksvareta},\ref{fig:gradmag}.
	
	\subsubsection{Fireball}
	\label{sec:ResFireball}
	For the fireball evolution (Figs. \ref{fig:fireball},\ref{fig:fireball2}) we find that since the value of $\eta$ is small at early times, the saturation radius initially resides below the photosphere, and both the temperature and luminosity increase significantly as a function of time.  After a few tens of seconds, $R_s>R_{\rm ph}$ and this trend is reversed; the luminosity and temperature gradually decrease as they begin tracking the same properties at the base of the jet. 
	
	Both the luminosity and temperature evolve dramatically, by between 2.5 and 5.5 orders of magnitude if the jet base is located near the NS surface, and between 3 and 6 orders of magnitude if the jet based is located at the light cylinder.  We are thus motivated to define the epoch at which the GRB emission starts, $t_0$, as the time at which the luminosity first reaches 10\% of its maximum value; this typically occurs $t_0=20-50$ s after the proto-NS formation.
	
	The late-time luminosity decay is controlled by the spin-down of the magnetar $\dot{E} \propto t^{-2}$ and is therefore too shallow as compared with observed rates during the ``early steep decay" phase of GRB afterglows. We find that for $B_{\rm dip,0}=2\times 10^{15}$ G the energy released in the first $\sim 100$ s (source frame) is $\sim 5-6\%$ of the rotational energy, depending on the assumption on the location of the base of the jet.  This value reaches $\sim 24-28\%$ for $B_{\rm dip,0}=10^{16}$ G. Hereafter, we will refer to this fraction of the magnetar initial rotational energy that is radiated away as "radiative efficiency of the jet".
	
	\subsubsection{Internal Shocks}
	\label{sec:ResIntShocks}
	Initially the photospheric radius may be larger than the shocks radius, resulting in optically thick emission, somewhat similar to the fireball model discussed above.  However, this phase is usually over before the jet breaks out of the star, or at most $\sim$ 10 s after the break out for the $B_{\rm dip,0}=2\times 10^{15}$G model (In addition, given the low luminosity and peak energy at these early times, this phase would likely not be observable even in this case).
	
	Although the normalizations of $E_{\rm peak}$ and $L_{\rm rad}$ depend on the unknown parameters $\epsilon_e,\epsilon_B,\xi$, their temporal evolution can be deduced assuming these unknown values do not evolve strongly in time.  As shown in Fig. \ref{fig:intshocks}, $L_{\rm rad}$  increases by roughly one order of magnitude, and $E_{\rm peak}$ by roughly four orders of magnitude, in the first $\sim 100$ s of the GRB. After this time, following the rapidly rising evolution of $\eta(t)$ the peak energy abruptly rises by several orders of magnitude.  Given the resulting high photon energies after this point, significant pair creation (not included in our calculation) will occur and reprocesses the high energy photons to softer emission. 
	
	Furthermore, around the same time that the jet undergoes the high $\sigma_0$ transition, we expect that its mass will become dominated by electron-positron pairs instead of baryons.  Since the energy per electron abruptly then decreases by a factor $\sim 1000$, the resulting synchrotron peak would drop by a factor $\sim 10^6$, to frequencies below the X-ray band.\footnote{However, note that this result is dependent on the pair multiplicity of the wind, which may be substantially lower for magnetars than for normal pulsars (e.g.~\citealt{Beloborodov13})}  Observationally, this high-$\sigma_0$ transition would likely mark the end of the GRB in the internal shock model, though prompt emission may continue in the X-ray band.  Taken together, one predicts strong temporal evolution in this model within a given burst, and a roughly fixed duration of $\sim 100$ s. Note however, that since the internal shocks model is characterised by a rather large emission radius ($\gtrsim 10^{15}$cm), the observed decay at the end of the prompt phase is dominated in this model by high-lattitude emission (\cite{KP2000}; see discussion in \S \ref{ESD}) rather than by the intrinsic change in luminosity.
	
	We also consider a variation of the shock model in Fig. \ref{fig:intshocksvareta} in which we introduce short timescale variability on top of the secular wind evolution (see \S \ref{sec:intShock}).  The time evolution in this case is much more sporadic in the first 10$-$30 s, during which the shocks are dominated by the internal variations rather than by collisions between the freshly ejected material and the bulk shell. Because of the large variation in luminosity within the variability timescale $\Delta t=1$ s, we show results for $E_p$ only at times when the luminosity exceeds 10\% of its maximum.  The lower limit on the emission radius of $\sim \eta(t)^2 c \Delta t$, ensures that the average luminosity at this time is almost constant and that $E_p$ is initially larger than in the previous case.  Despite the modifications introduced by stochastic variability, we still predict time-variation in $E_{\rm p}$ of order $\sim 2$ orders of magnitude during this phase (inconsistent with observations).  For $B_{\rm dip,0}=2\times 10^{15}$ G, the radiated energy during the first $\sim 100$ s is only $\sim 2\%$ of the magnetar's initial rotational energy in this variation of the internal shocks model, as compared with $\sim 1.5\%$ for the model with no random variability in $\eta$ (or $\sim 7\%$ as compared to $\sim 6\%$ for $B_{\rm dip,0}=10^{16}$ G). These values are calculated for $\epsilon_e=1$, but smaller $\epsilon_e$ would lead to even lower radiative efficiencies.
	
	\subsubsection{Gradual Magnetic Dissiption}
	\label{sec:ResGradMag}
	The spectrum in gradual magnetic dissipation models is quasi-thermal at early times due to the low initial magnetization of the wind.  However, given the delay time $t_{\rm bo} \sim 10$ s required for the jet to escape from the star following magnetar formation, this completely thermal phase of the burst may be unobservable, or at most observable only for the first few seconds near the beginning of the GRB.  Between $t \sim 10$ s and $30-100$ s (see Fig. \ref{fig:gradmag}) the peak energy (which is dominated by the thermal component) decreases moderately from $\sim 70$ to $\sim 20$keV, mainly due to the increasing value of $\lambda \propto \Omega^{-1}$ as the proto-NS spins down.  During the same time interval, the low energy (10-50 keV) spectra gradually becomes softer, due to the increase in the relative strength of the optically thin to photospheric component of the emission as $\eta(t)$ grows. 
	
	In the next several tens of seconds, the increasing value of $\eta(t)$ eventually overcomes the decreasing value of  $L(t)$ and $\Omega(t)$ to cause a rise of the peak energy.  This growth in $\eta(t)$ also leads to a slow-cooling spectrum (\S \ref{sec:magdiss}), and thus to an overall significant reduction of the radiated luminosity over a very brief interval of time.  The net result is a strong upper limit on the GRB duration in this model of $T_{90} \lesssim 200$ s, after which the GRB would become too faint to detect.  One implication of this is that, unless the spin-down time of the GRB is less than $200$ s (or equivalently $B_{\rm dip,0}\gtrsim 3\times 10^{15} I_{45}^{1/2}P_{0,-3}R_6^3$ G), the total GRB energy is again significantly less than the rotational energy of the magnetar.  The radiative efficiency for $\epsilon_e=0.2$ is $1.5\%$ for $B_{\rm dip,0}=2\times 10^{15}$ G  (or $\sim 7\%$ for $B_{\rm dip,0}=10^{16}$ G).  The maximum efficiency ($\epsilon_e=1$) reaches $3\%$ for $B_{\rm dip,0}=2\times 10^{15}$ G  (or $16\%$ for $B_{\rm dip,0}=10^{16}$ G).
	
	\subsection{Effect of beaming angle}
	\label{sec:beaming}
	Our fiducial models assume a canonical value for the beaming fraction of $f_b=1.9\times 10^{-3}$ (jet opening angle of $\theta_j=5^{\circ}$).  In this section we explore the dependence of our results on the value of $f_{b}$.
	
	\subsubsection{Fireball}
\begin{figure*}
		\centering
		\includegraphics[scale=0.35]{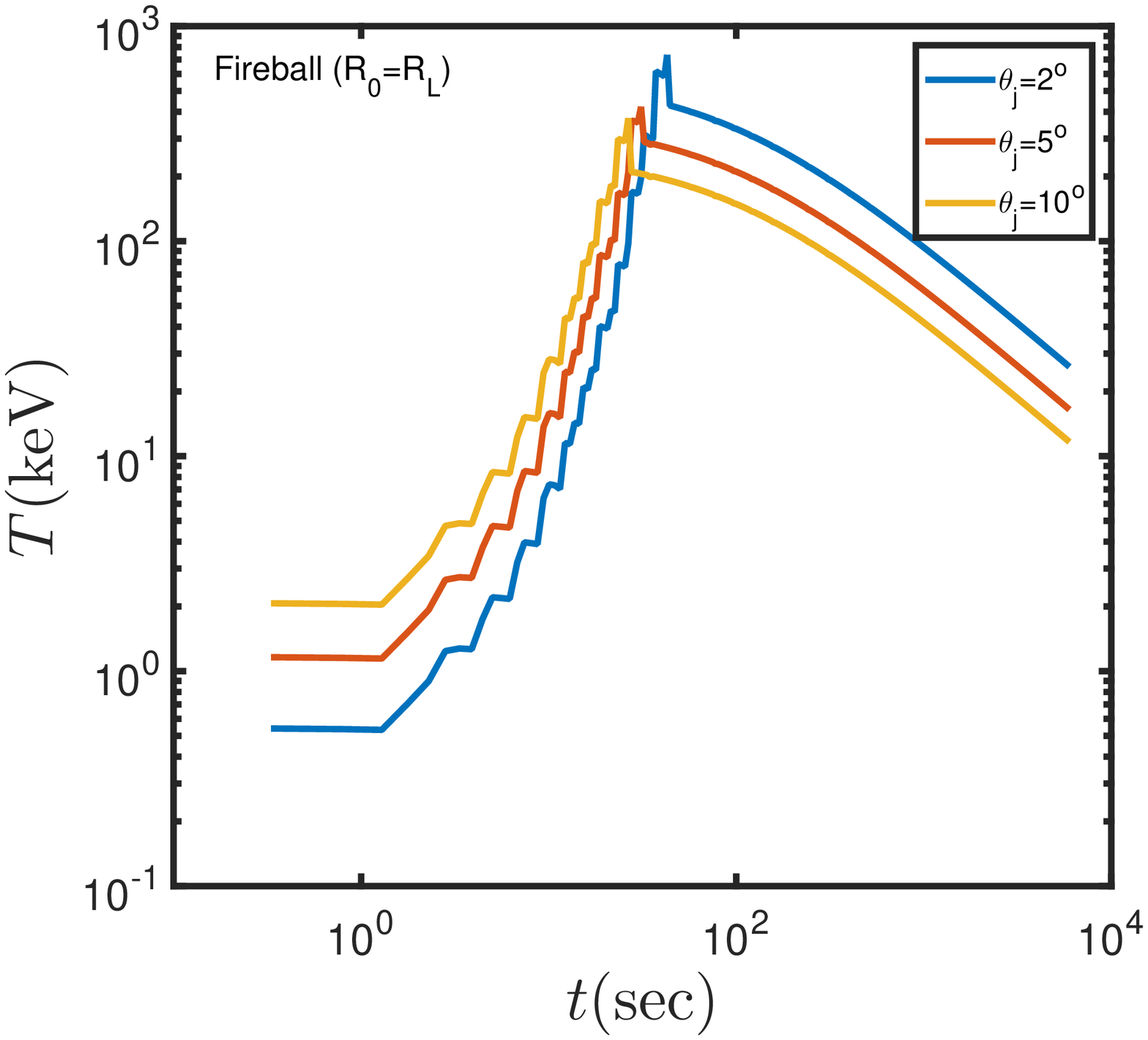}
		\includegraphics[scale=0.35]{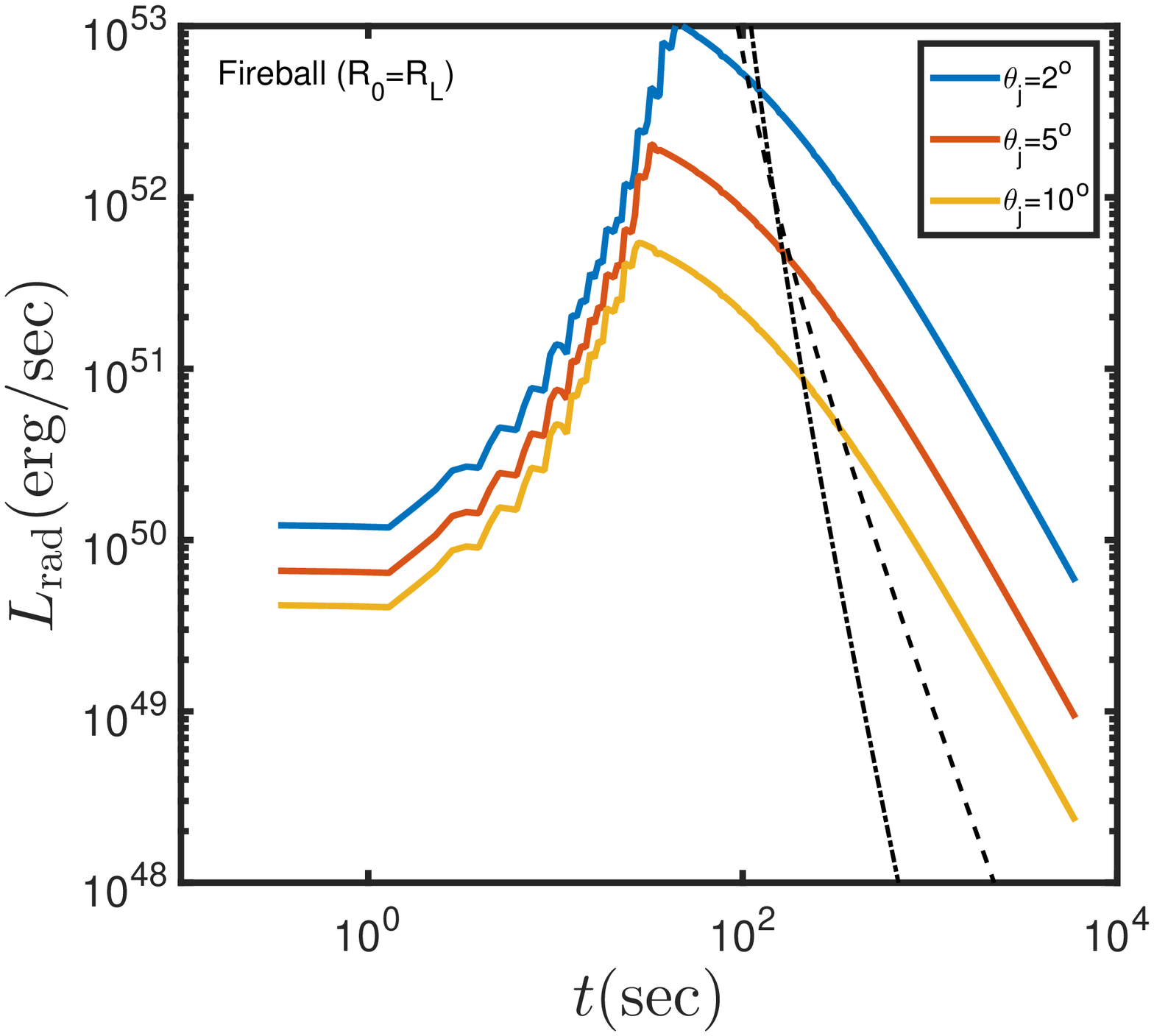}
		\caption
		{\small Evolution of observed temperature (left) and isotropic equivalent luminosity (right) of a fireball driven by a magnetar central engine, for different jet opening angles. Results are shown for $B_{\rm dip,0}=10^{16}$ G, assuming the base of the jet is at the light cylinder, assuming a breakout time $t_{\rm bo}=10$s and a redshift $z=1$. Time is measured in the observer frame starting from the breakout time. On the RHS, dashed and dot-dashed lines mark $(t-t_0)^{-3}$ and $(t-t_0)^{-5}$ decay rates respectively (typical to the observed early steep decays, as described in Fig. \ref{fig:fireball}).}
		\label{fig:fireballtheta}
	\end{figure*}

	Fig. \ref{fig:fireballtheta} shows the temporal evolutions of luminosity and temperature for the fireball model with varying values of the opening angle. For clarity, we only show here results for the case of a constant $B_{\rm dip,0}=10^{16}$ G with the jet base located at the light cylinder. 
	
	The main parameter affected by $f_b$ is the asymptotic luminosity. The general dependence of the observed temperature and luminosity is similar in all cases and can be understood as follows. The temperature at the base of the jet (which is also the observed temperature when $R_s>R_{\rm ph}$) scales as $T_0\propto f_b^{-1/4}$ (see Eq. \ref{eq:T0}), whereas the corresponding isotropic equivalent luminosity scales as $L_{0,\rm iso}\propto f_b^{-1}$. These conditions hold once $\eta$ becomes larger than $\eta_*$, which typically occurs a few tens of seconds after the onset of the burst (see \S \ref{sec:Fireball}). For $R_s<R_{\rm ph}$, since the location of the photosphere scales as $f_b^{-1}$ (see Eq. \ref{eq:photo}) we obtain $L_{\rm iso}\propto L_{0,\rm iso} (R_{\rm ph}/R_0)^{-2/3}\propto f_b^{-1/3}$, $T\propto T_0 (R_{\rm ph}/R_0)^{-2/3}\propto f_b^{5/12}$.  Overall, the peak luminosity in the fireball model scales approximately as $L_{\rm iso}=2\times 10^{52} (\theta_{j}/5^{\circ})^{-2}\mbox{erg s}^{-1}$.\footnote{The actual scaling with $\theta$ is slightly shallower because for larger $\theta$, $R_{ph}$ becomes larger than $R_s$ earlier, when $\dot{E}$ is somewhat larger.}

	\subsubsection{Internal Shocks}
\begin{figure*}
		\centering
		\includegraphics[scale=0.35]{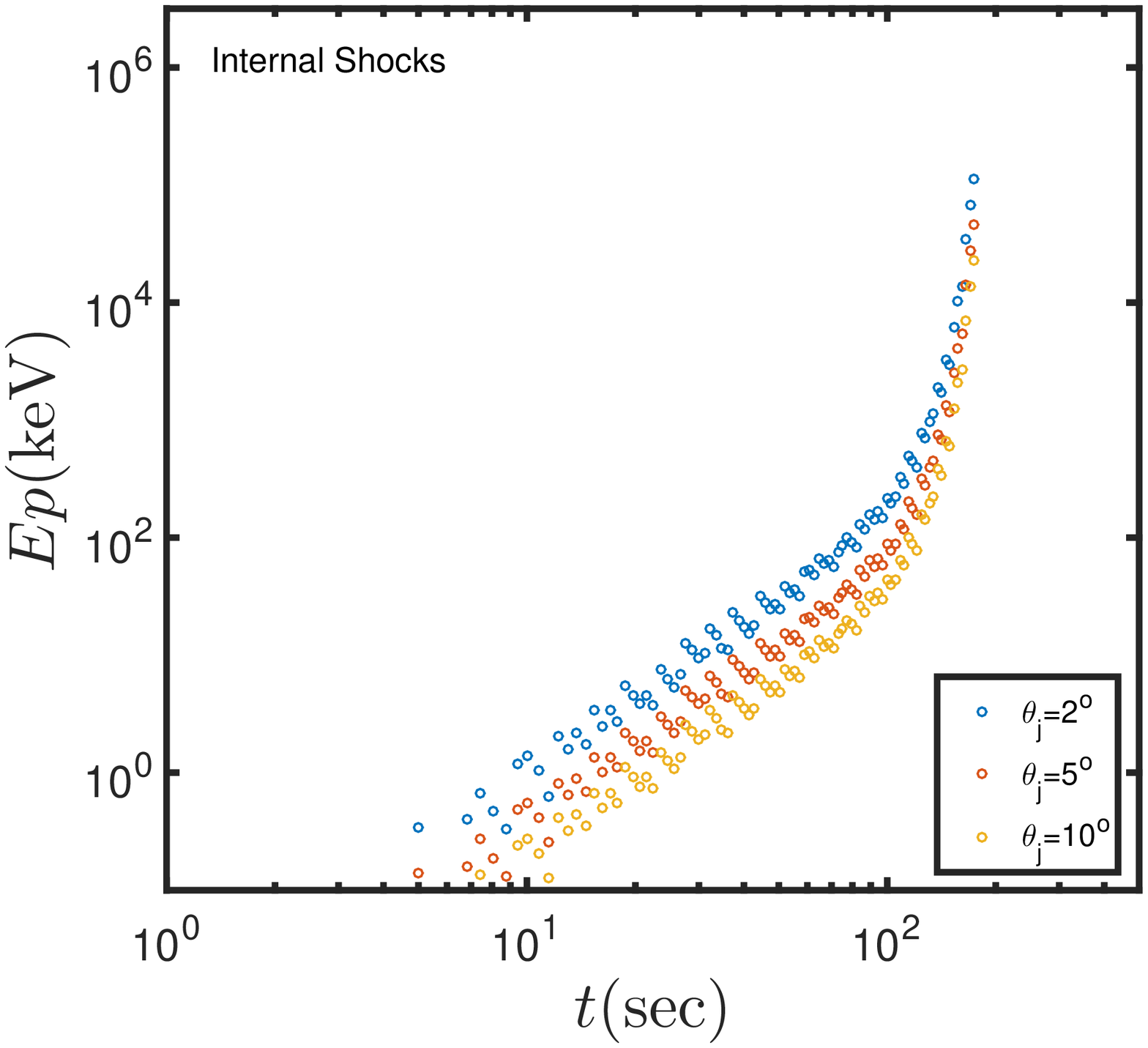}
		\includegraphics[scale=0.35]{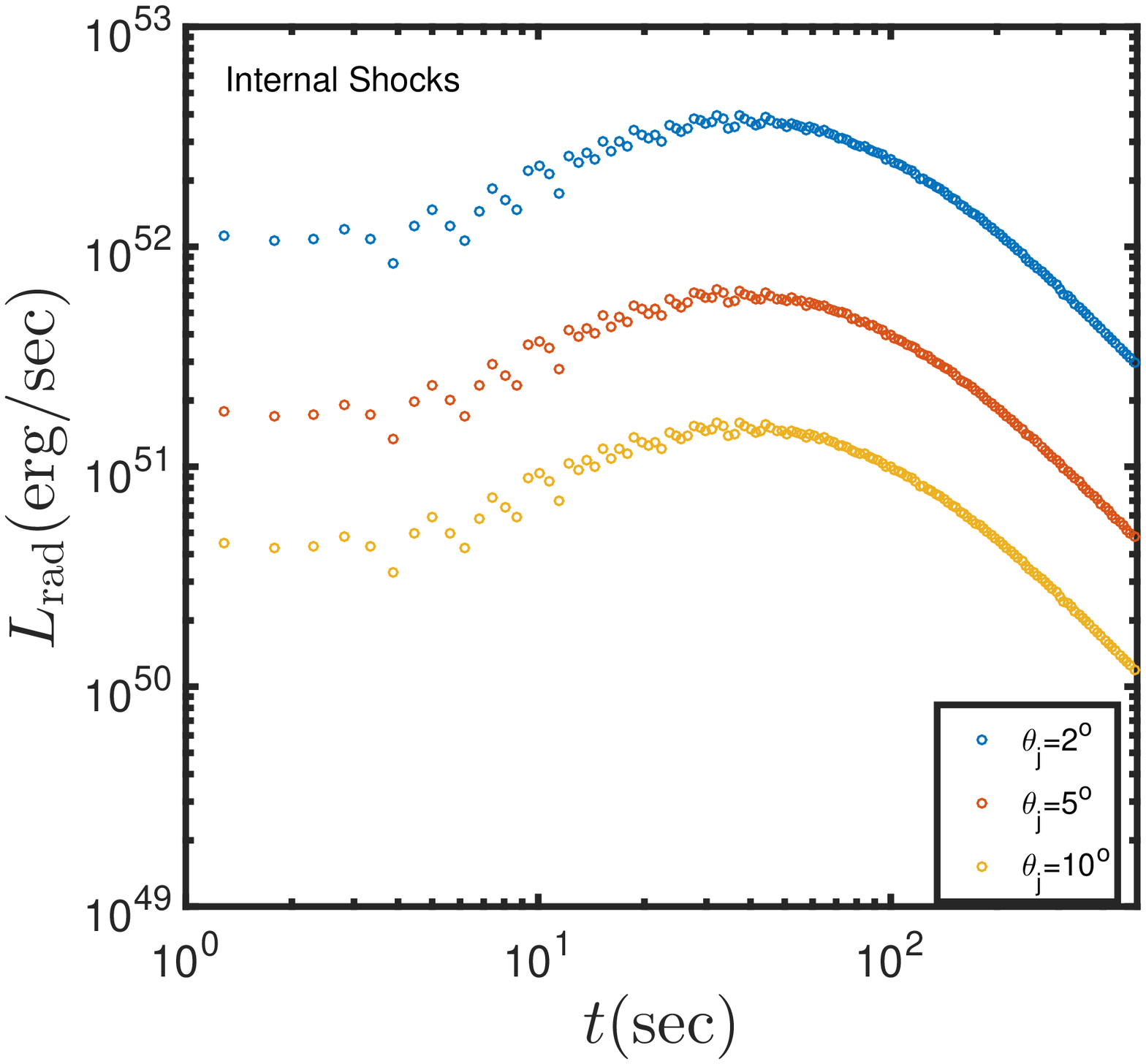}
		\caption
		{\small Evolution of observed temperature (left) and isotropic equivalent luminosity (right) of internal shocks originating from a magnetar central engine, for different jet opening angles. Results are shown for $B_{\rm dip,0}=10^{16}$ G, assuming a breakout time $t_{\rm bo}=10$s and a redshift $z=1$. Time is measured in the observer frame starting from the breakout time. We also assume here: $\epsilon_e=1,\epsilon_B=0.01,\xi=0.01$.}
		\label{fig:intshockstheta}
	\end{figure*}

	Fig. \ref{fig:intshockstheta}  shows the affect of varying the jet opening angle on the temporal evolutions of luminosity and temperature in the internal shocks model.  For clarity, we only show results for the case without random variations in $\eta$, as the scalings with $f_b$ are the same in both cases.  The shapes of these curves are independent of the beaming and as can be seen by Eq. B13 of \cite{Metzger+11}: the isotropic equivalent luminosity and peak energy scale as $L_{\rm iso} \propto f_b^{-1}$ and $E_p\propto f_b^{-1/2}$, respectively.  The peak luminosity scales as $L_{\rm iso}=6\times 10^{51} (\theta_{j}/5^{\circ})^{-2}\mbox{erg s}^{-1}$.  Note that even for an optimistic case of $\theta_j=2^{\circ}$, the peak luminosity is only $\sim 4\times 10^{51}\mbox{erg s}^{-1}$.

	\subsubsection{Gradual Magnetic Dissipation} 
\begin{figure*}
		\centering
		\includegraphics[scale=0.35]{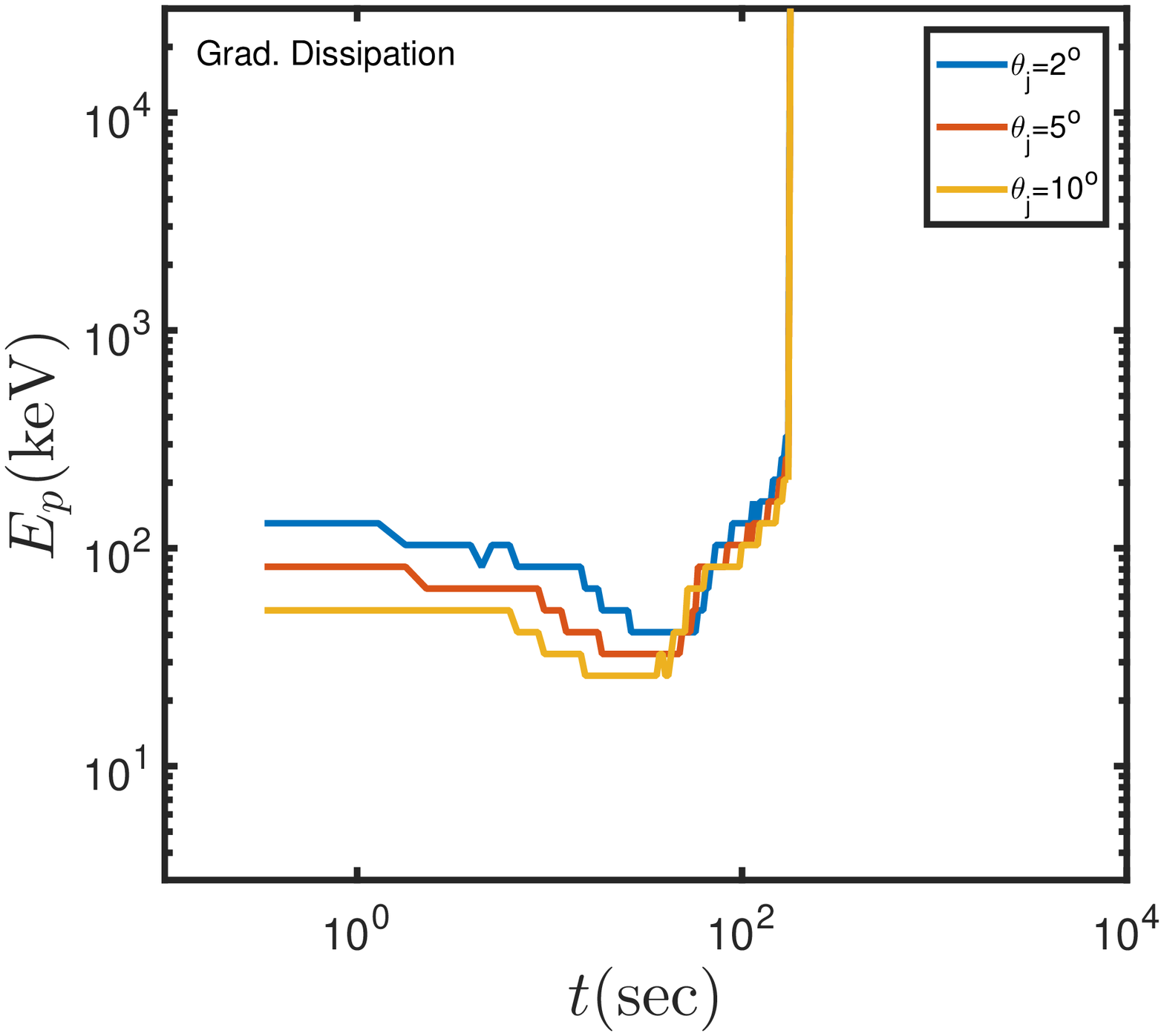}
		\includegraphics[scale=0.35]{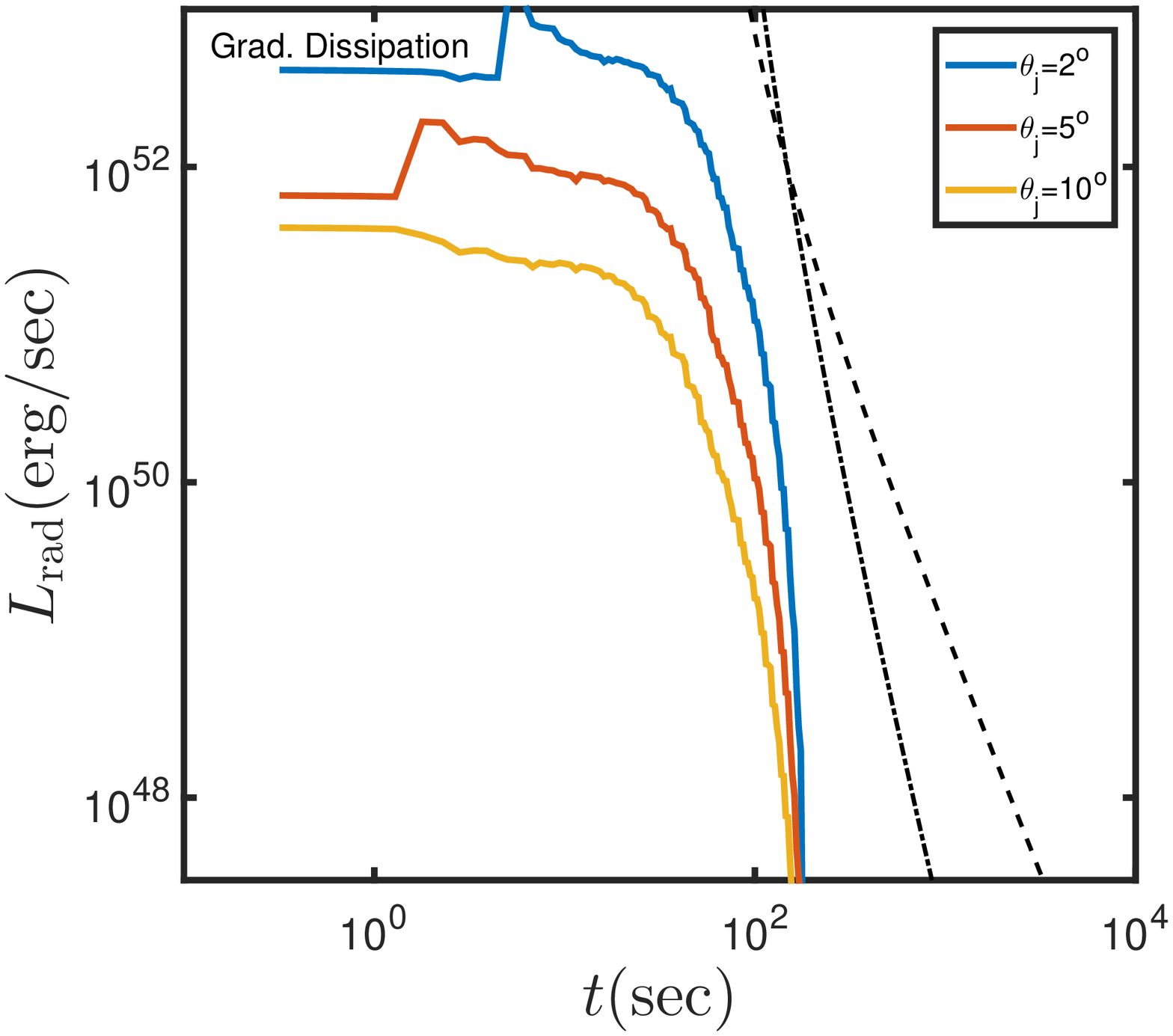}
		\caption
		{\small Evolution of observed temperature (left) and isotropic equivalent luminosity (right) for a jet originating from a magnetar central engine and undergoing gradual magnetic dissipation, for different jet opening angles. Results are shown for $B=10^{16}$ G, assuming a breakout time $t_{\rm bo}=10$s and a redshift $z=1$. Time is measured in the observer frame starting from the breakout time. We also assume here: $\epsilon_e=0.2,\xi=0.2,\epsilon=0.2$. On the RHS, dashed and dot-dashed lines mark $(t-t_0)^{-3}$ and $(t-t_0)^{-5}$ decay rates respectively (typical to the observed early steep decays, as described in Fig. \ref{fig:fireball}).}
		\label{fig:gradmagtheta}
	\end{figure*}

Fig. \ref{fig:gradmagtheta} shows the temporal evolutions of luminosity and temperature for the gradual magnetic reconnection model with varying values of the opening angle. The shapes of these curves are similar in all cases (\citealt{BG2017}, their eqns. 16, 17), assuming that the spectral peak is dominated by the thermal component, the isotropic equivalent luminosity and the peak energy approximately scale as $L_{\rm iso} \propto f_b^{-6/5}$ (roughly $L_{\rm iso}=2\times 10^{52} (\theta_{j}/5^{\circ})^{-12/5}\mbox{erg s}^{-1}$), $E_p\propto f_b^{-1/10}$. These scalings are not exact since both parameters are also affected to some extent by the synchrotron component, for which the different characteristic frequencies (and corresponding fluxes) have different beaming dependencies. The most important of these are the peak frequency and flux in the fast cooling regime, $\nu_p\propto f_b^{1/2}$, $\nu L_{\nu}(\nu_p)\propto f_b^{-6/5}$.

	\section{Comparison to GRB Observables}
	\label{sec:General}
	This section compiles predictions of the magnetar model for a range of observable properties of GRB prompt emission, which follow from the detailed analysis given in the previous sections:
	\subsection{Prompt Emission Duration}
	As shown in \S \ref{sec:fiducmodel}, the predicted GRB duration is $T_{90}\sim 100$ s in both the internal shocks and gradual magnetic dissipation models.  In both cases, this is set by the timescale $t_{\sigma_0}$ at which $\sigma_0$ abruptly increases rapidly as the proto-NS becomes optically thin.  In internal shock models, the cut off at high $\sigma_{0}$ is the result of significant pair creation in the jet following the rapid increase of the peak emission energy, while in the magnetic dissipation model the prompt emission ceases due to reduced radiative efficiency at high $\sigma_{0}$ as the emitting electrons become slow cooling.  In the fireball scenario, the burst trigger in the observer frame may be delayed compared to the onset of magnetar formation by $t_0\approx 20-50$ s due to the luminosity initially being too low to detect.  The end of the burst is set by the spin-down timescale of the magnetar 
	\begin{equation}
	\label{eq:tSD}
	t_{\rm SD}\!\approx \! 25\,{\rm s} \,\,\!\bigg(\!\frac{M}{1.4M_{\odot}}\!\bigg)\! \bigg(\frac{B_{\rm dip,0}}{ 10^{16}\mbox{G}}\bigg)^{-2}\! \bigg(\frac{P_0}{1.5\mbox{ms}}\bigg)^2 \bigg(\frac{R_{\rm NS}}{12 \mbox{km}}\bigg)^{-4},
	\end{equation}
	which for $B_{\rm dip,0}=2\times10^{15}-10^{16}$ G is typically tens to hundreds of seconds.
	
	Both $t_{\rm SD}$ and $t_{\sigma_0}$ depend on the initial magnetar parameters, as summarized in Table \ref{tbl:NSparams} for different sets of initial parameters.  Predicted GRB durations are in the range of tens to hundreds of seconds in all models, consistent with the observed $T_{90}$ distribution.
	
	We note that the required surface magnetic dipole field strength may be somewhat smaller than those estimated here, because we have assumed that the transition radius between open and closed magnetic field lines, $R_Y$ (the so-called Y-point radius), occurs at the light cylinder, $R_L$.  However, relativistic MHD simulations suggest that for relatively low magnetization, $R_{\rm Y}$ is smaller than $R_L$ (e.g.~$R_{\rm Y}\approx 0.3 R_L$ for $\sigma_0 \approx 1$, and scaling weakly with $\sigma_0$ as $\sigma_0$ rises; \citealt{Bucciantini2006}).  In this situation, the same magnetic flux, and thus magnetar spin-down luminosity, could be obtained for a smaller value of the surface magnetic field.  Since the open magnetic flux $\Phi$ which enters the force-free spin-down rate scales as $\Phi \propto B_{\rm dip,0} R_Y^{-1}$, a reduction of $R_Y$ by a factor 3 would imply that the same spin-down luminosity and time-scales could be achieved for a surface field which is also 3 times weaker.  Note, however, that $t_{\sigma_0}$ would be largely unaffected by this change.
	
	A shorter GRB duration could be obtained if the proto-magnetar were to suddenly collapse to a black hole at time $t_{\rm collapse} ,< T_{90}$. However this premature demise will also reduce the overall radiative efficiency (discussed in the next point) even further, by a factor $\sim t_{\rm collapse}/T_{90}$. Thus shorter GRBs in the magnetar model are expected to have similar luminosities to those of longer GRBs, but radiated energies that are smaller by a factor of order the duration ratio, as observed \citep{Ghirlanda2009}.  
	
	At the other end of the duration distribution, ultra long GRBs, with $T_{90}\approx 10^4$ s \citep{Levan2014} have also recently been suggested to originate from magnetar central engines \citep{Metzger2015,Gompertz2017}, motivated in part by the recent coincidence of one such event with a very luminous supernova \citep{Greiner+15}.  It is less clear how magnetar emission models described here could accommodate for such large time-scales, especially if the duration is dominated by $t_{\sigma_0}$ (which, as discussed below, is the favored scenario given other observed properties).
	
	\begin{table*}
		\begin{center}
			\caption{Energetics and timescales for a characteristic range of proto-magnetar outflow models.}
			\scalebox{0.75}{
				\begin{tabular}{cccccccc} \hline \hline
					$B$ (G) & $P_0$(ms) & $M_{\rm NS} (M_{\odot})$ & $\chi$  & $t_{\rm SD}$(s)$^1$& $t_{\sigma_0}$(s)$^2$ & $E_{tot} (t_{\rm SD}) (10^{50} \mbox{erg})^{3}$ & $E_{\sigma_0} (10^{50} \mbox{erg})^3$\\ \hline 
					$2 \times 10^{15}$ & 1.5 & 1.4 & $\pi/2$ & 771 & 74 & 25 & 2.3\\
					$2 \times 10^{15}$ & 1.5 & 2 & $\pi/2$ & 1040 & 122 & 39 & 5\\
					$2 \times 10^{15}$ & 1.5 & 1.4 & $0$ & 1540 & 63 & 25 & 1.2 \\
					$2 \times 10^{15}$ & 1.5 & 2 & $0$ & 2080 & 109 & 39 & 2.6 \\
					$2 \times 10^{15}$ & 1 & 1.4 & $\pi/2$ & 330 & 70 & 58 & 11\\
					$2 \times 10^{15}$ & 1 & 2 & $\pi/2$ & 440 & 109 & 93 & 23 \\
					$10^{16}$ & 1.5 & 1.4 & $\pi/2$ & 38 & 41 & 25 & 13 \\
					$10^{16}$ & 1.5 & 2 & $\pi/2$ & 46 & 64 & 39 & 24 \\
					$10^{16}$ & 1.5 & 1.4 & $0$ & 67 & 31 & 25 & 11 \\
					$10^{16}$ & 1.5 & 2 & $0$ & 86 & 46 & 39 & 18\\
					$10^{16}$ & 1 & 1.4 & $\pi/2$ & 23 & 39 & 58 & 30 \\
					$10^{16}$ & 1 & 2 & $\pi/2$ & 26 & 58 & 93 & 51\\
					\hline  \\
					\multicolumn{8}{c}{1 time at which the rotational energy reaches half its original value.}\\
					\multicolumn{8}{c}{2 time at which $\sigma_0=10^4$.}\\
					\multicolumn{8}{c}{3 Energy released by the magnetar wind prior to $t_{\rm SD}$ ($0.5E_{\rm rot}$).}\\
					\multicolumn{8}{c}{4 The total energy released by the magnetar wind between $\sigma_0=100$ and $\sigma_0=3000$.}\\
					
					\label{tbl:NSparams}
				\end{tabular}
			}
			
		\end{center}
	\end{table*}
	
	\subsection{Energetics} The total available energy reservoir of a magnetar is its initial rotational energy
	\begin{equation}
	E_{\rm rot}\approx 1.5\times 10^{52}\mbox{erg} \bigg(\frac{M}{1.4M_{\odot}}\bigg)\bigg(\frac{R_{\rm NS}}{12 \mbox{km}}\bigg)^2 \bigg(\frac{P_0}{1.5\mbox{ms}}\bigg)^{-2}.
	\end{equation}
	For a $1.4M_{\odot}$ NS, spinning at roughly the break-up velocity, $P_0=1$ ms, $E_{\rm rot}\approx 3\times 10^{52}$erg. This energy is already low compared to lower limits of the total (collimated corrected) radiated energy in some GRBs \citep{Beniamini2015,Beniamini2016}. For more massive NSs, with $M=2M_{\odot}$, $P_0$ can conceivably be as small as $P_0=0.7$ ms \citep{Metzger2015}, allowing for larger values of $E_{\rm rot}\approx 10^{53}$ erg. The total energies released by the wind within $t_{\rm SD}$ and between $t(\sigma_0=100)$ and $t_{\sigma_0}$ for different magnetar parameters, are shown in Fig.~\ref{fig:dEdsigma} and listed in Table \ref{tbl:NSparams}; typical values are in the range of $ 10^{50}-5\times 10^{51}$ erg.
	
	The total energy radiated in the gamma-ray band is reduced by the efficiency of dissipation and radiation as well as by the fact that the spin down time-scale may either be considerably longer than the GRB duration, resulting in too little energy release before $t_{\sigma_0}$ or alternatively be so short that a significant fraction of the energy is released before $\sigma_0=100$ (see \S \ref{sec:magnetarevolve} and Fig. \ref{fig:dEdsigma}).  The range of models we have considered result in radiative efficiencies of $\sim  1-28\%$ relative to the entire rotational energy budget during the first 100 s of the emission.  The highest efficiencies require fireball-like evolution with $B_{\rm dip,0} = 10^{16}$G, but more typical values are $\sim 5\%$.
	
\begin{figure*}
		\centering
		\includegraphics[scale=0.35]{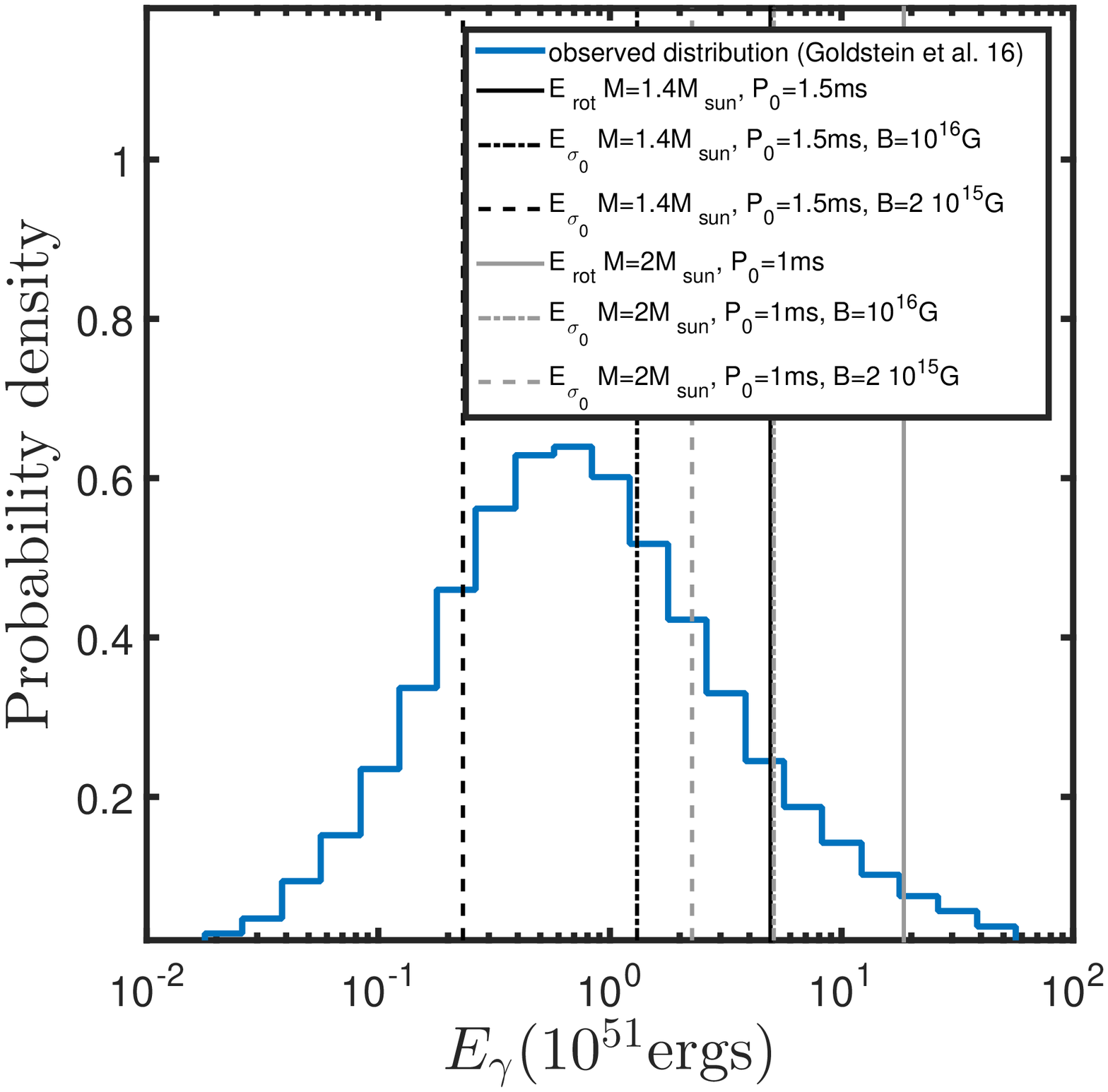}
		\includegraphics[scale=0.35]{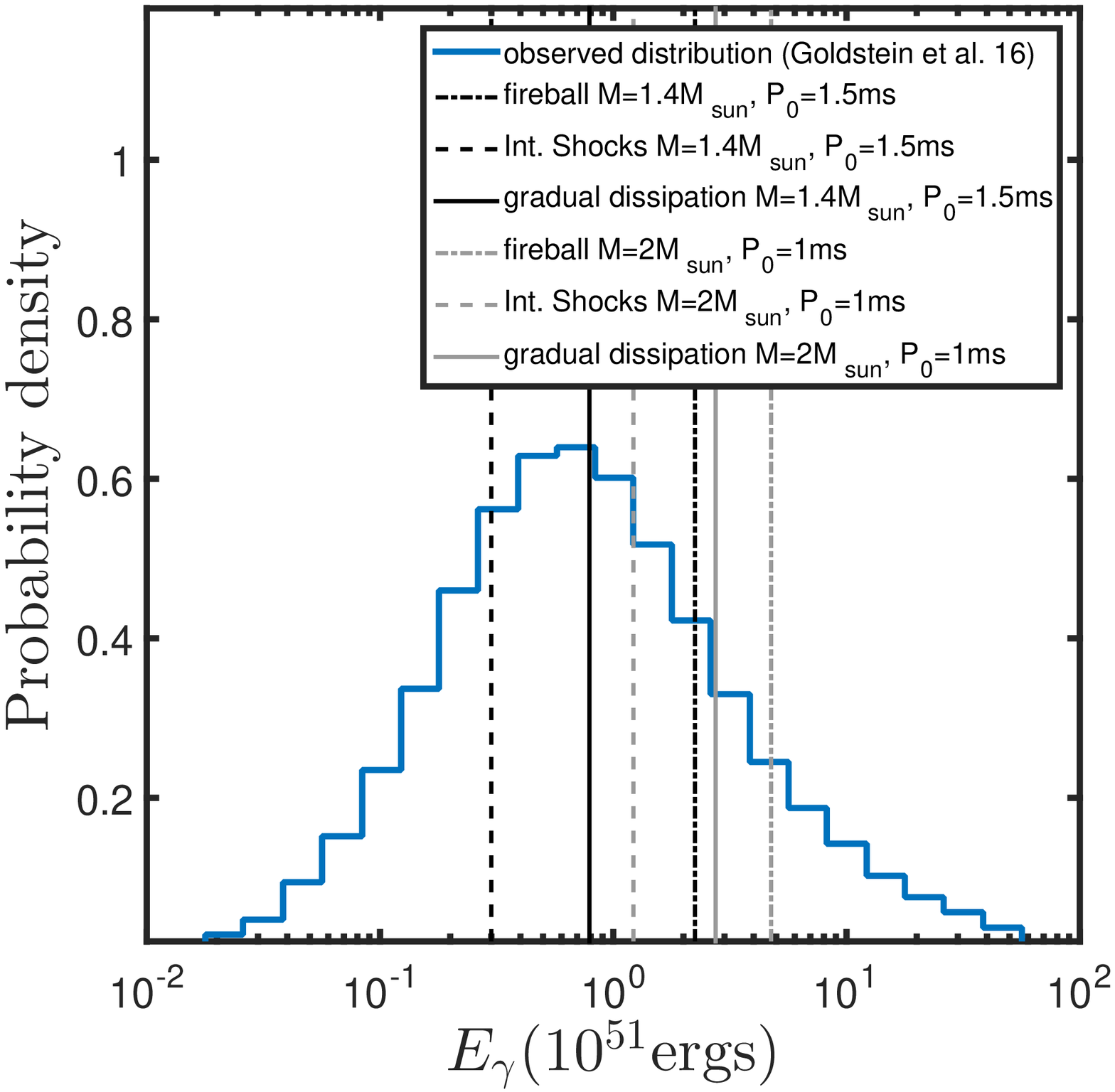}
		\caption
		{\small {\bf Left:} Comparison between the observed distribution of GRB radiated $\gamma$-ray energies from  \citet{Goldstein2016} with various energetic constraints on the magnetar model, including the total rotational energy as well as the wind energy $E_{\sigma_0}$ which is released in the magnetization interval $\sigma_{0} = 100-3000$.
			{\bf Right:} Same as the left panel, but comparing the observed GRB distribution to the maximum radiated energy for different prompt emission models studied in this paper assuming a NS dipole field $B=10^{16}$G.  In both panels we have assumed a magnetar with inclination angle $\chi=\pi/2$, while we show separately results for a magnetar with mass 1.4$M_{\odot}$ NS and initial spin  $P_0=1.5$ ms as well as one with 2$M_{\odot}$ NS and $P_0=1$ ms.}
		\label{fig:Egamma}
	\end{figure*}

	Fig. \ref{fig:Egamma} compares rough limits on the collimated corrected energies within the models discussed here to a distribution of measured values of the collimated corrected energies found by \cite{Goldstein2016}. The observed distribution is symmetrical in log-space and peaks at $\sim 6\times 10^{50}$erg, whereas the highest $\gamma$-ray energy expected in the gradual the dissipation and internal shocks models is $2.5 \times 10^{51} / 1.3 \times 10^{51}$erg respectively, obtained for a NS with $2M_{\odot}$, $B_{\rm dip,0}=10^{16}$G and $P_0=1$ ms. As the observed distribution extends to values larger than the most optimistic case by at least 1.5 orders of magnitude, this appears to disfavor all GRBs originating from magnetar central engines. 
	
	\cite{Ryan2015} have shown that since GRBs are typically observed slightly off-axis, the actual opening angles (and hence also the collimated corrected energies), when this effect was not taken into account, have been overestimated. The difference between the average observed angle implied by \cite{Goldstein2016} and \cite{Ryan2015} is $\sim 27\%$ leading to a factor 2 effect on the $\gamma$-ray energies. This is still short of bridging the gap between expectations from the most massive and fast spinning magnetars and the observed distribution. An exception is the fireball model, that is able to emit large amounts of energy due to the fact that the GRB does not end at $t_{\sigma_0}$. However, this model is disfavored for other reasons, based on its predicted gamma-ray spectrum (see below).
	
	One caveat is that we have assumed the magnetar spins down in isolation following the supernova explosion.  In reality, it may still be accreting matter from the surrounding fall-back of the explosion in the equatorial plane defined by the rotation axis of the progenitor star (e.g., \citealt{Piro&Ott11,Bernardini+14}). Depending on the mass fall-back rate, this accretion could supply the magnetar with additional angular momentum (and thus rotational energy), allowing the total energy budget of the GRB jet to exceed that of the maximum rotational energy at any one time. In this case, the outflow geometry of the magnetar wind would be similar to that described by \cite{Parfrey+16} and the spin-down rate would be increased from the isolated force-free expression we have adopted due to the larger open magnetic flux.  However, we note that the maximum extractable rotational energy from the magnetar drops rapidly once the magnetar grows to a mass above the maximum stable mass $M_{\rm max} \sim 2.1-2.4M_{\odot}$ for a non-rotating neutron star, because in this case the magnetar needs to maintain significant rotational support just to avoid collapsing to a black hole (\citealt{Metzger2017}; see his Fig.~8). 

	\subsection{Temporal Evolution} As shown in \S \ref{sec:fiducmodel}, we expect strong temporal evolution of both the peak energy and luminosity in both the fireball and internal shocks models. Such rapid temporal evolution is not observed in most GRB prompt emission \citep{Ford1995,Lu2012}.  For the gradual magnetic dissipation model, by contrast, the evolution is comparatively moderate (at least until $t \approx 100$ s when $\sigma_{0}$ rapidly increases, terminating the GRB). This behaviour is not trivial and is the result of the evolution of the typical luminosity and peak energy in the gradual dissipation model with the evolution of $\dot{E}(t),\sigma_0(t),\Omega(t)$ implied by the magnetar model.  
	
	In order to explore the generality of this conclusion to other classes of gradual dissipation models, we consider a generalized model to that of \cite{BG2017}, where the flow accelerates as a power-law in radius $\Gamma \propto r^n$ (see also \citealt{Veres2013}), thus reducing to our fiducial model in the special case $n=1/3$.  In this case, the temperature and luminosity of the photosphere (which control the peak of the GRB emission in gradual dissipation models) depend on the jet parameters as follows,
	\begin{eqnarray}
	\label{TphLph}
	E_p\propto \dot{E}^{5n-1 \over 8n+4}\sigma_0^{8-19n \over 8n+4}\Omega^{7n \over 8n+4} \\
	L\propto \dot{E}^{1+3n \over 1+2n}\sigma_0^{-5n \over 1+2n}\Omega^{n \over 1+2n}
	\end{eqnarray}
	which gives $E_p \propto \sigma_0^{3/4}\Omega^{1/4}$ for $n=1/5$ or $E_p \propto \dot{E}^{1/3}\sigma_0^{-11/12}\Omega^{7/12}$ for $n=1$. 
	
\begin{figure*}
		\centering
		\includegraphics[scale=0.35]{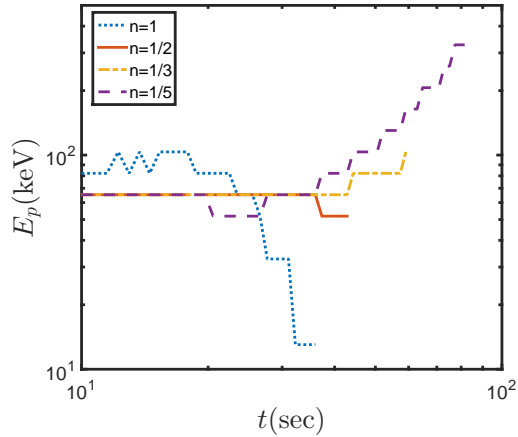}
		\caption
		{\small Evolution of the peak GRB energy for generalized magnetic dissipation models with varying acceleration rates $\Gamma \propto r^n$ and a magnetar central engine. Results are shown for $B_{\rm dip,0}=10^{16}$ G, assuming a breakout time $t_{\rm bo}=10$s and a redshift $z=1$.}
		\label{fig:Epeakdiffn}
	\end{figure*}

	The peak emission in gradual dissipation models depends most sensitively on $\sigma_0$, which is also the most rapidly-evolving jet property during the burst in the magnetar model.  As $\sigma_0(t)$ increases during the burst, the peak energy will increase for $n \lesssim 0.4$ or decrease for $n\gtrsim 0.4$.  Fig. \ref{fig:Epeakdiffn} shows the result of inserting the specific evolution of $\dot{E}(t),\sigma_0(t),\Omega(t)$ for a magnetar with a constant dipole field of $10^{16}$ G for different values of $n$.  We only show $E_p$ at those times when the magnetar luminosity still exceeds 10$\%$ of its maximum value during the initial phase because beyond this point a typical GRB will fall below the detectability threshold. Even with this restriction, we find that $1/5 \lesssim n \lesssim 1$ is required for the peak energy to vary by $\lesssim 10$ during the observed duration of the GRB.  We thus conclude that values of $n \approx 1/3$ similar to those predicted for the magnetic dissipation model are required to reconcile the magnetar model with the observed evolution of GRB prompt emission.
	
	\subsection{Spectrum} The spectrum of prompt GRB emission is non-thermal (\citealt{Band1993}), which historically has led to the suggestion of synchrotron emission as the dominant radiation process (\citealt{Katz1994,Rees&Meszaros94,Sari1996,Kumar2008,Daigne2011,Beniamini2013}). In synchrotron-dominated models for emission in the fast cooling regime (expected for instance for the internal shocks model discussed here), the spectrum below the Band peak has a power-law with spectral index $\alpha=-1.5$ (where $dN/d\nu \propto \nu^{\alpha}$) or softer. This is in strong tension with typical observed values of $\alpha=-1$ and harder values observed in many GRB (a.k.a. the ``death line"; \citealt{Crider1997,Preece1998,Preece2000}).  Furthermore, several studies have discovered evolution of the low energy spectrum (in the $\sim$ 10-50 keV range) from hard to soft during the burst \citep{Kargatis1994,Ford1995,Band1997}.
	
	For the pure fireball model, at times when the saturation radius formally lies above the photosphere, the emission is efficient and the spectrum will closely resemble a blackbody.  As such it will be incompatible with the observed spectra, in particular predicting a low energy spectrum which is much too hard ($\alpha\approx +1$, however slightly softer spectra are obtained if the outflow energy is still accelerating at the photosphere, or $R_s>R_{\rm ph}$ \cite{Beloborodov2011}). Furthermore, as the initial temperature is predicted to be lower than 10 keV, we would expect an initial strong hardening of the 10-50 keV spectrum as the temperature crosses through this band and until the spectral index saturates at the value above. 
	
	For internal shocks, the initial value of the peak energy depends on the unknown values of $\epsilon_e,\epsilon_B,\xi$.  If the peak energy resides below 10 keV, one would predict an initial hardening of the 10$-$50 keV spectrum as the peak energy crosses through the observed band, after which the low energy spectral index saturates at a maximum of $\alpha=-1.5$, corresponding to fast cooling synchrotron. Alternatively, if the peak exceeds 50 keV at early times, then one should observe $\alpha=-1.5$ throughout the burst's evolution.  Both the fireball and internal shocks scenarios are clearly incompatible with the observed spectral properties of GRBs. 
	
	By contrast, for the gradual magnetic dissipation model, the spectrum is a combination of photospheric and optically thin (non-thermal) emissions.  It can typically be fit by an intermediate spectrum with $\alpha$ between -1 and 0, consistent with the measured low energy spectral slope.  In addition, the low energy spectrum in the magnetar model initially softens with time (as observed; see above) until $\sim 30-100$ s after the burst (when the burst becomes highly inefficient and $\alpha$ increases significantly).  The bottom right panel of Fig. \ref{fig:gradmag} shows the evolution of $\alpha$ for the gradual magnetic dissipation model

	\subsection{Variability}
	
	For both fireball and gradual magnetic dissipation models, the dynamical variability, $R/2c \Gamma^2$ is considerably smaller than the typically observed variability in GRBs of $\approx$ 1 s (\citealt{Fishman1995,Norris1996,Quilligan2002}). A possible candidate for the variability timescale, the spin period of the magnetar, is typically milliseconds, much shorter than this timescale.  However, long $\sim 1$ second-timescale variability may be naturally explained from jet propagation through the star \citep{Morsony2010,Bromberg2016,LopezCamara2016}, in which case it would be independent of the engine model.
	
	For the internal shocks model without random variations in $\eta(t)$, the emission radius is sufficiently large, such that $R/2c\Gamma^2$ is indeed in the range favored by observations. However, as shown by \cite{Metzger+11}, this quantity increases roughly linearly in time, contrary to GRB observations which show no systematic change in the variability timescale throughout the GRB (\citealt{RamirezRuiz2000}). Introducing random variations in $\eta(t)$ by hand, as discussed in \S \ref{sec:intShock}, would maintain the variability time near a constant value of $\Delta t\sim 1$ s for the first 10$-$30 s of the burst (depending on the magnetic field strength), after which time the emission radius is no longer controlled by these internal variations and the predicted secular rise in the variability time discussed above would be recovered.
	
	\subsection{Early Steep Decay}
	\label{ESD}
	Observations of the early afterglows of GRBs reveal a ``steep decay" phase \citep{Tagliaferri2005} during which the flux decays approximately as a power-law with a decay index of $3-5$. The observed decay rate in most GRBs is compatible with the predictions for high-latitude emission \citep{KP2000}, which is the geometrical viewing effect originating from the fact that photons emitted at larger angles take longer to reach the observer, so even if the emission at the source shuts off very rapidly, a tail of flux still persists to later times. In order for the high-latitude effect to dominate the observed decay, the angular time-scale, $t_{\theta}=R/2c \Gamma^2$ must be of the order of seconds. This constraint is generally satisfied for the internal shocks model, but not so in the other two models considered here. In both the fireball and gradual magnetic dissipation models, the typical emission radius is $\lesssim 10^{13}$cm, resulting in an angular time-scale which is too short.  For both of these cases, the early steep decay must instead be dominated by the actual decay of the engine luminosity. 
	
	For the fireball emission model, the engine decay is dominated by that of the spin-down luminosity, $\dot{E} \propto t^{-2}$, which is too shallow to be consistent with observations.  This remains true even when taking into account the fact that the prompt emission trigger may be 20$-$50 s after the burst, due to the burst luminosity initially being too low to detect in fireball models (see Figs. \ref{fig:fireball},\ref{fig:fireball2}, right panels). For the gradual dissipation model, the luminosity shuts off too rapidly due to the reduction in efficiency as $\eta$ increases (see Fig. \ref{fig:gradmag}, right panel). We note however, that as mentioned in \S \ref{sec:magnetarevolve}, at late times, the evolution of $\sigma_0$ may become shallower due to even moderate entrainment from the jet walls.  Extra baryon loading at late times could therefore alleviate the tension with the observed steep decay phase within the gradual magnetic dissipation model. 
	
	\subsection{Luminosity - peak energy correlation} Various studies have found the existence of a correlation between the peak luminosities and peak energies of GRBs such that $L_p \propto E_p^{1.5-2}$ \citep{Yonetoku2004,Ghirlanda2004}. A test of the viability of these relations within the magnetar model, requires knowledge of the unknown distribution of the intrinsic magnetars' properties, such as their masses, their magnetic fields, their initial spins etc. and therefore cannot be done at this  stage. However, recently it was also suggested that similar relations between the luminosity and peak energy hold also within a specific GRB, when comparing different pulses and isolating the non-thermal part of the GRB emission \citep{Guiriec2015A,Guiriec2015B}. These studies typically find that $L_{\rm NT}=(9.6 \pm 1.1) (E_{p,\rm source, NT}/100\mbox{keV})^{1.38\pm 0.04}\mbox{erg s}^{-1}$.
	
	As seen in Figs. \ref{fig:intshocks}, \ref{fig:intshocksvareta} the typical trends between these parameters in the internal shocks model are significantly shallower than the observed correlation, approximately satisfying $L_p\propto E_p^{1/4-1/3}$ (as mentioned in \S \ref{sec:intShock} the normalization in this case can easily be changed by adopting different values for $\epsilon_e,\epsilon_B, \xi$). For the gradual magnetic dissipation model, we find that the resulting trend is somewhat closer to the observed one (see Fig. \ref{fig:gradmag}). It should also be stressed that to obtain the observed correlation, one needs to assume a specific spectral model for the non-thermal part. This has typically taken to be Band or Band + PL, which is different than the shape of the non-thermal spectrum in the gradual dissipation model. It is therefore reasonable to expect some difference between the observed and theoretical correlations in this case.
	
	\subsection{Delayed onset of GRB emission}
	In the gradual dissipation model, no temporal gap between break-out and the GRB is expected. However, in both the internal shocks and fireball model, the peak energies are initially low and the GRB may only be detected a few tens of seconds after the jet break-out. Such a gap could be observed if the jet break-out emission can be detected due its own emission \citep{RamirezRuiz2002,Waxman2003,Zhang2003,Nakar2012}. Another source of early time emission could be IC scattering of photons emitted by the hot cocoon surrounding the GRB jet on relativistic electrons in the jet \citep{Kumar2014}. Observationally, about $\sim 20\%$ of long GRBs \citep{Lazzati2005} and $\sim 10\%$ of short GRBs \citep{Troja2010} are found to have had a weaker precursor in $\gamma$-rays occurring between a few seconds to a few tens of seconds before the GRB trigger.
	
	\subsection{Plateaus}
	Many GRBs exhibit extended X-ray plateaus, lasting up to tens of thousands of seconds after the GRB trigger. Various authors have considered a magnetar origin as the source of energy powering these plateaus (e.g. \citealt{Zhang2006,Liang2006,Troja2007}; see however \citealt{Uhm2007,Genet2007,BM2017}). In this scenario, the spin-down time and luminosity of the magnetar are associated with the duration and luminosity, respectively, of the plateau. However, if magnetar spin-down accounts for the plateau, then it is unclear what sets the duration and energetics of the prompt emission.  Clearly, in such a scenario, insufficient rotational energy would have been released by the magnetar during the much shorter period of the prompt GRB.
	
	The plateau duration (luminosity) is typically three to four orders of magnitude longer (smaller) than that of the prompt phase, implying a very different required values for the magnetar parameters.  Given the relations for the spin-down luminosity $L_{\rm SD} \propto \Omega^4 B_{\rm dip,0}^2$ and timescale $t_{\rm SD} \propto B_{\rm dip,0}^{-2}\Omega^{-2}$, the main constraint is on the magnetic field, which is required to be $\sim 10^{14}$G in order to be able to account for X-ray plateaus, i.e. $1.5-2$ orders of magnitude smaller than the values required by the prompt emission in our models.
	
	A possible way out of this apparent contradiction, which still maintains a magnetar origin for both phases, is to invoke decay of the magnetic field, such that initially $B_{\rm dip} \approx 10^{16}$ G (corresponding to $t_{SD} \sim 100$ s) and then the field decays to $B_{\rm dip} \sim 3\times 10^{14} G$ with a new spin-down time $t_{\rm SD} \gtrsim 10^{4}$ s.  However, this possibility seems somewhat contrived, and requires fine-tuning in order for the field decay to happen by a time that is roughly comparable to the initial spin-down time, i.e. the prompt GRB duration (if the field decays much earlier, the prompt would be significantly underpowered as compared to the plateau and vice versa if the field decays much later). Amongst the models we have considered in this work, the convection equipartition model for the magnetic field, has the right qualitative behaviour. However, assuming the field saturates at its value just prior to the end of the neutrino emission phase, the total change in $B_{dip,0}$ by a factor of $\sim 4$ is insufficient to account for the required transition from typical prompt time-scales and luminosities to the typical values of those properties during the plateau phase.
	
	Another possibility is that the plateau is powered by late-time accretion onto the central engine. Fall-back accretion may dominate the magnetar spin-down between up to hours after the collapse, which is a typical time-scale for X-ray plateaus. It remains however to be determined whether this scenario could realistically result in a constant value of $\dot{E}$, which is large enough to account for the observed luminosities, and yet does not lead to a collapse of the magnetar to a black hole. 
	\section{Conclusions}
	\label{sec:conclusions}
	We summarize our main conclusions below.
	\begin{itemize}
		\item The magnetar model provides quantitative estimates for the evolution of the intrinsic parameters at the base of the GRB wind such as the energy injection and energy per baryon. This implies that magnetar engines have more specific predictions (which may naturally lead them to also become more constrained) than BH engines. 
		\item The most important parameter controlling the GRB evolution in the magnetar model is $\sigma_0$. GRB emission is typically restricted to $t(\sigma_0=100)-t_{\sigma_0}$ where the lower limit results from compactness constraints and the upper limit ($t_{\sigma_0}\approx t(\sigma_0=3000)$) marks the rapid rise in magnetization over a brief period of time when the neutrino emission luminosity from the proto-magnetar drops approaching the magnetar's transition to optically thin conditions. Considering different spin periods and dipole magnetic fields, the amount of energy released during this time is at most 0.25 of the initial rotational energy, even before accounting for any further potential losses due to the dissipation or emission mechanisms.
		
		The rapid rise in $\sigma_0$, at $t_{\sigma_0}$, translates into strong expected spectral evolution in most models, and implies a strong limit on possible GRB durations.  Magnetar models are challenged to explain the very long durations of ultra-long GRBs, unless the jet entrains significant baryonic mass from the surrounding jet walls.
		\item Different dissipation models for the GRB prompt emission predict a radiative efficiency of $\approx 1-28\%$. and thus a total (collimated corrected) energy of order $\lesssim 5\times 10^{51}$erg, even for the optimistic case of very massive (2$M_{\odot}$) and rapidly spinning ($P_0=1$ ms) magnetars. Taking the observed energy distribution at face value, this means that at most $\sim 70\%$ of long GRB could be accounted for by the magnetar model (Fig.~\ref{fig:Egamma}).  At the same time, however, this also demonstrates how magnetar models naturally produce the observed energy scale of typical GRBs. 
		
		The situation should be compared with black holes that are the alternative for magnetar central engines. Energy-wise they have no limit, which alleviates the maximal energy problem. However, it is perhaps surprising that the energy distribution does peak at $\sim 10^{51}$erg in this case as this requires accreting only $10^{-2}M_{\odot}$ at 10$\%$ jet efficiency of $\dot{M}c^2$. This is part of a broader issue with black holes, which is that they do not offer robust predictions for the evolution of $\sigma_0$ (see, however, \citealt{Lei2013}) or any other of the model parameters. This is currently the main source of uncertainty in determining the identity of the central engine.
		\item Considering specific dissipation models, we find that both the internal shocks and pure fireball models for magnetar central engines are in strong contention with GRB observations. The main issue being that these models predict a strong evolution of the luminosity and peak energy as a function of time that is not observed.

		By contrast, the gradual magnetic dissipation results in fairly `realistic' GRBs in terms of many key observables, such as energy, duration, spectrum and the lack of strong temporal evolution during the burst. This agreement results simply by substituting the predicted time evolution of the magnetar wind parameters into the predictions of the gradual dissipation model, with relatively little fine-tuning required.  Similar agreement extends to a (limited) class of photospheric models, which distribute the jet heating relatively uniformly across a range of radii in the jet (Fig.~\ref{fig:Epeakdiffn}). 
		
		\item Our model has focused on jets powered by the spin-down evolution of an isolated proto-magnetar, which neglects the effect of fall-back accretion from the supernova explosion (as occurs typically on timescales of minutes to hours after the collapse for stripped-envelope GRB progenitor stars). Fall-back in sufficient quantity can add mass and angular momentum to the magnetar, altering its spin-down evolution and changing the predicted properties of the GRB emission. Additional mass accretion could in principle result in a premature collapse of the magnetar to a black hole, resulting in a shorter gamma-ray burst duration and a smaller amount of released energy. On the other hand, the additional angular momentum from the fall-back could keep the magnetar spinning rapidly longer than it would in isolation, thus enhancing the total rotational energy available for powering the burst. The potentially complex evolution of an accreting proto-magnetar, and its effect on the prompt GRB emission properties, will be explored in future work.
		
	\end{itemize}
	\section*{acknowledgments}
	DG acknowledges support from NASA through the grants NNX16AB32G and NNX17AG21G issued through the Astrophysics Theory Program.  BDM acknowledges support from NSF through the grant AST-1410950, and from NASA through the Astrophysics Theory Program grant NNX16AB30G and Swift Guest Investigator Program grant NNX16AN77G.
	
	\bibliographystyle{mnras}
	\bibliography{magnetar}
\end{document}